\documentclass[12pt,pre,aps,showpacs,preprint]{revtex4}

\usepackage{graphicx}
\usepackage{amsfonts}
\usepackage{amssymb}
\usepackage{amsmath}
\usepackage{float}

\begin{document}

\title{ Reformulation of the Covering and Quantizer Problems
as Ground States of Interacting Particles}

\author{S. Torquato}

\email{torquato@electron.princeton.edu}

\affiliation{Department of Chemistry, Princeton University,
Princeton New Jersey 08544, USA}

\affiliation{Department of Physics, Princeton University,
Princeton New Jersey 08544, USA}

\affiliation{Princeton Center for Theoretical Science, Princeton
University, Princeton New Jersey 08544, USA}

\affiliation{Program in Applied and Computational Mathematics,
Princeton University, Princeton New Jersey 08544, USA}

\affiliation{Princeton Institute for the Science and Technology of
Materials, Princeton University, Princeton New Jersey 08544, USA}

\date{\today}

\begin{abstract}
It is  known that the sphere packing problem and the number variance
problem (closely related to an optimization
 problem in  number theory) can be posed as energy 
minimizations associated with  an infinite number 
of point particles in $d$-dimensional Euclidean space $\mathbb{R}^d$ interacting
via certain repulsive pair potentials.  We reformulate the covering and quantizer 
problems as the determination of the ground states of interacting particles in  $\mathbb{R}^d$
that generally involve single-body, two-body,
three-body, and  higher-body interactions. This is done by linking  
the covering and quantizer problems to  certain optimization problems
involving the ``void" nearest-neighbor functions that 
arise in the theory of random  media and statistical mechanics. These reformulations,
which again exemplifies the deep interplay between geometry and physics,
allow one now to employ theoretical and numerical optimization techniques 
to analyze and solve these energy minimization problems. 
The covering and quantizer problems
have relevance in numerous applications, including wireless communication network layouts, the
search of high-dimensional data parameter spaces, 
stereotactic radiation therapy, data compression, digital communications, meshing of space for numerical analysis, and 
coding and cryptography, among other examples.
In the first three space dimensions, the best known solutions of the sphere packing and number
variance problems (or their ``dual" solutions) are directly related to those of the covering and quantizer problems, but such  relationships 
may or may not  exist for $d \ge 4$, depending on the peculiarities
of the dimensions involved.  Our reformulation sheds light
on the reasons for these similarities and differences.
We also show that disordered saturated sphere packings provide relatively thin (economical) coverings
and may yield thinner coverings than the best known lattice coverings
in sufficiently large dimensions.
In the case of the quantizer problem, 
we derive improved upper bounds on  the quantizer error using sphere-packing solutions, 
which  are generally substantially 
sharper than an existing upper bound in low to moderately large dimensions.
We also demonstrate that {\it disordered}  saturated sphere packings yield
relatively good quantizers. Finally, we remark on possible applications of our
results for the detection of gravitational waves.

\end{abstract}

\maketitle

\section{Introduction}

There are certain scientific problems that provide deep connections
between many different scientific fields. The study of   the low-energy states of 
classical interacting many-particle systems is an exemplar of a class of
such problems because of its manifest importance in physics, materials science, communication theory, cryptography, mathematics and computer science. Such many-particle systems
have been used with great success to model liquids, glasses
and crystals when quantum effects are negligible \cite{Ha86,To02a}.
The total potential energy $\Phi_N({\bf r}^N)$ of $N$ identical
particles with positions ${\bf r}^N \equiv {\bf r}_1,{\bf r}_2,\ldots,{\bf r}_N$
in some large volume in $d$-dimensional Euclidean space $\mathbb{R}^d$
can be resolved into separate one-body, two-body, $\ldots$ , $N$-body contributions:
\begin{equation}
\Phi_N({\bf r}^N)=\sum_{i=1}^N u_1({\bf r}_i) + \sum_{i<j}^N u_2({\bf r}_i,{\bf r}_j) +  \sum_{i<j<k}^N u_3({\bf r}_i,{\bf r}_j,{\bf r}_k) + \cdots + u_N({\bf r}^N),
\label{full}
\end{equation} 
where $u_n$ represents the intrinsic $n$-body interaction in excess
to the interaction energy  for $n-1$ particles. To make the statistical-mechanical
problem more tractable, the exact many-body
potential (\ref{full}) is usually replaced by a mathematically simpler form.
For example, in the absence of an external field (i.e., $u_1=0$),
often one assumes pairwise additivity, i.e.,
\begin{equation}
\Phi_N({\bf r}^N)=\sum_{i<j}^N u_2({\bf r}_i,{\bf r}_j).
\label{pair} 
\end{equation}
Pairwise additivity is exact for hard-sphere systems
and frequently has served to approximate accurately
the interactions in simple liquids, such as the well-known Lennard-Jones
pair potential \cite{Ha86}, and
in more complex systems where the pair potential 
$u_2$ in (\ref{pair}) can be regarded to
be an {\it effective} pair interaction \cite{St02}.

An outstanding problem in classical statistical mechanics
is the determination of the {\it ground states} of $\Phi_N({\bf r}^N)$,
which are those configurations that globally minimize $\Phi_N({\bf r}^N)$
and hence are the states that exist at absolute zero temperature.
While classical ground states are readily produced by
slowly freezing liquids in experiments and computer simulations,
our theoretical understanding of them is far from
complete \cite{Ru99,To08a}. Virtually all theoretical/computational ground-state studies 
of many-particle systems have 
been conducted for pairwise additive potentials \cite{Ha86,Chaik95,Ru99,La00,St01,Ml06,Co07,Gr08}.
Often the ground states of short-range pairwise interactions
are crystal structures in low dimensions \cite{Ha86,Chaik95,Ru99,La00,St01,Ml06,Co07,Gl07},
but long-range interactions exist that can suppress
any kind of symmetry leading to disordered ground states
in low dimensions \cite{Uc04b,Ba08}. Moreover, in
sufficiently high dimensions, it has been suggested
that even short-ranged pairwise interactions possess
disordered ground states \cite{To06a,To06b,Sc08}.

Ground states of purely repulsive pair interactions have 
profound connections not only to low-temperature states
of matter but to problems in pure mathematics, including discrete
geometry and number theory \cite{Co93,Sa06,Co07}, information
theory, and computer science.
As will be explained further below, it is  known that the sphere packing problem and the number variance problem (closely related to an optimization
 problem in  number theory) can be posed as energy 
minimizations associated with  an infinite number 
of point particles in $d$-dimensional Euclidean space $\mathbb{R}^d$ interacting
via certain repulsive pair potentials. Both of these problems
can be interpreted to be optimization problems involving {\it point processes},
which can then be recast as energy minimizations.
A point process in $\mathbb{R}^d$ is a distribution of an an infinite number of points in $\mathbb{R}^d$ at number density $\rho$ (number of points per unit volume) with configuration ${\bf r}_1,{\bf r}_2,\ldots$; see Ref. \cite{To06b}
for a precise mathematical definition. 

A packing of congruent nonoverlapping spheres is a special point process in which there is a minimal pair separation distance,
equal to the sphere diameter.
The sphere packing problem seeks to determine the densest arrangement(s)
of congruent, nonoverlapping $d$-dimensional spheres in Euclidean space $\mathbb{R}^d$ \cite{Co93,Co03}.
Although it is simple to state, it is a notoriously difficult problem
to solve rigorously. Indeed, Kepler's four-century-old
conjecture, which states  that the face-centered-cubic lattice in $\mathbb{R}^3$
is maximally dense, was only recently proved \cite{Ha05}. For $d \ge 4$, the packing problem remains unsolved \cite{Co93,Co03,Co09}.
It is well known that the sphere packing problem can be posed as an energy 
minimization problem  involving pairwise interactions between 
points in $\mathbb{R}^d$ (e.g., inverse power-law functions in which the exponent tends to infinity);
see Ref.~\cite{Co07} and references therein.

Problems concerning the properties and quantification of density fluctuations in 
many-particle systems  continue to provide many theoretical challenges.
Of particular interest are  density fluctuations that occur on some 
local length scale \cite{To03a}. It has been shown that the  minimal number variance
associated with points (e.g., centroids of atomic or molecular systems) contained within some ``window" can also
be formulated as a ground-state problem involving bounded
repulsive pair interactions with compact support \cite{To03a,support}.
For spherical windows in the large-radius limit, the best known solutions in 
$\mathbb{R}^d$ 
are usually point configurations that are ``duals" (in the sense
discussed later in the paper) to the best known
sphere packings in $\mathbb{R}^d$.

The focus of this paper is on two other optimization problems involving
point processes in $\mathbb{R}^d$: the {\it covering} and {\it quantizer} problems.
Roughly speaking, the covering problem asks for the point configuration
that  minimizes the radius of overlapping spheres circumscribed around 
each of the points required to cover $\mathbb{R}^d$. The covering problem
has applications in wireless communication network layouts \cite{Ad02}, the
search of high-dimensional data parameter spaces (e.g.,
search templates for gravitational waves) \cite{Pr07}, and stereotactic radiation therapy \cite{Li09}. The quantizer problem 
is concerned with finding the point configuration in $\mathbb{R}^d$ that
minimizes a ``distance error" associated with a randomly placed point
and the nearest point of the point process. It has applications in computer science
(e.g., data compression) \cite{Co93}, digital communications \cite{Co93},
coding and cryptography \cite{Bo02}, and optimal meshing of space
for numerical applications (e.g., quadrature and discretizing
partial differential equations) \cite{Du99}.
Heretofore, the covering and quantizers problems were not
known to correspond to any ground-state problems.

We reformulate the covering and quantizer 
problems as the determination of the ground states of interacting particles in  
$\mathbb{R}^d$ that generally involve single-body, two-body,
three-body, and higher-body interactions. This is done by linking  
the covering and quantizer problems to  certain optimization
problems involving the ``void" nearest-neighbor functions that 
arise in the theory of random  media and statistical mechanics \cite{To02a,Re59,To90c}. These reformulations,
which again exemplifies the deep interplay between geometry and physics, enable one to employ  theoretical and numerical optimization techniques 
to solve these energy minimization problems. We find that disordered {\it saturated} sphere packings 
(roughly, packings in which no space exists to add an additional
sphere) provide relatively thin (i.e., economical) coverings
and may yield thinner coverings than the best known lattice coverings
in sufficiently large dimensions. In the case of the quantizer problem, 
we derive improved upper bounds on  the quantizer error that utilize
sphere-packing solutions. These improved bounds are generally substantially 
sharper than an existing upper bound in low to moderately large dimensions.
We also demonstrate that disordered  saturated sphere packings yield
relatively good quantizers. 
Our reformulation helps to explain why the known solutions of quantizer and covering problems
are identical in the first three space dimensions and why they can be 
different for $d \ge 4$. In the first three space dimensions, the best known solutions of the sphere packing and number variance problems are directly related to those of the covering and quantizer problems, but such relationships  may or may not  exist for $d \ge 4$, depending on the peculiarities of the dimensions involved.

We begin by summarizing basic definitions and concepts in Sec.~\ref{def}.
Because of the connections between the sphere-packing,
number-variance, covering and quantizer problems, in Sec.~\ref{problems}, we 
formally define each of these problems, summarize
key developments, and compare
the best known solutions for each of them in selected
dimensions. This includes calculations obtained for the best
known number-variance solutions for $d=12,16$ and 24. We then define in Sec.~\ref{near} the void nearest-neighbor functions 
and represent them in terms of series involving certain
integrals over the $n$-particle correlation
functions, which statistically characterize an ensemble
of interacting points. The special case of a single
realization of the point distribution follows
from this ensemble formulation, which reveals that
quantizer and covering problems can be expressed
as ground-state solutions of many-body interactions
of the general form (\ref{full}). Section~\ref{reform} specifically
gives these ground-state reformulations and shows
how some known solutions in low dimensions
can be explicitly recovered using the void nearest-neighbor functions.
In Sec.~\ref{results-cover}, we show that {\it disordered} saturated sphere packings provide relatively thin coverings
and may yield thinner coverings than the best known lattice coverings
in sufficiently large dimensions. 
In Sec.~\ref{results-quant}, 
we derive improved upper bounds on  the quantizer error that utilize
sphere-packing solutions. 
We also show that {\it disordered}  saturated sphere packings yield
relatively good quantizers. Finally, in Sec.~\ref{conc},
we make concluding remarks and comment on the application
of the quantizer problem to the search for gravitational waves.

\section{Definitions and Preliminaries}
\label{def}

For a statistically homogeneous point process in $\mathbb{R}^d$
at number density $\rho$ (number of points per unit volume), the quantity
$\rho^n g_{n}({\bf r}_1,{\bf r}_2,\ldots, {\bf r}_n)$  is proportional to
the probability density for simultaneously finding $n$ sphere centers at
locations ${\bf r}_1,{\bf r}_2,\dots,{\bf r}_n$ in $\mathbb{R}^d$ \cite{Ha86}.
 With this convention, each {\it $n$-particle correlation function} $g_n$ approaches
unity when all of the points become widely separated from one another.
Statistical homogeneity implies that $g_n$ is translationally
invariant and therefore only depends on the relative displacements
of the positions with respect to some arbitrarily chosen origin of the system, {\it i.e.},
\begin{equation}
g_n=g_n({\bf r}_{12}, {\bf r}_{13}, \ldots, {\bf r}_{1n}),
\end{equation}
where ${\bf r}_{ij}={\bf r}_j - {\bf r}_i$. As we will see,
statistically homogeneous point processes include as special
cases periodic point distributions.

The {\it pair correlation} function $g_2({\bf r})$ is a particularly
important quantity. If the
point process is also rotationally invariant (statistically
isotropic), then $g_2$ depends on the radial distance 
$r \equiv |{\bf r}|$ only, {\it i.e.}, $g_2({\bf r}) = g_2(r)$. 
Thus, it follows that the expected number
of points $Z(R)$ found in a sphere of radius $R$ from a randomly
chosen point of the point process, called the {\it cumulative
coordination function}, is given by
\begin{equation}
Z(R)=\rho \int_0^R s_1(r) g_2(r) dr,
\label{cum}
\end{equation}
where 
\begin{eqnarray}
s_1(r)  =  \frac{2\pi^{d/2}r^{d-1}}{\Gamma(d/2)}
\label{area-sph}
\end{eqnarray}
is the surface area of a $d$-dimensional sphere of radius $r$.   

A {\it lattice} $\Lambda$ in $\mathbb{R}^d$ is a subgroup
consisting of the integer linear combinations of vectors that constitute a basis for $\mathbb{R}^d$
and thus represents  a special subset of point processes.
In a lattice $\Lambda$, the space $\mathbb{R}^d$ can be geometrically divided into identical regions $F$ called {\it fundamental cells}, each of which contains 
the just one point  specified by the {\it lattice vector}
\begin{equation}
{\bf p}= n_1 {\bf a}_1+ n_2 {\bf a}_2+ \cdots + n_{d-1} {\bf a}_{d-1}+n_d {\bf a}_d,
\end{equation}
where ${\bf a}_i$ are the basis vectors for the fundamental cell 
and $n_i$ spans all the integers for $i=1,2,\cdots d$. We denote by 
$v_F$  the volume of the fundamental cell. In the physical sciences, a lattice is equivalent
to a Bravais lattice. Unless otherwise stated, we will
use the term lattice. Every lattice has a dual (or reciprocal) lattice $\Lambda^*$
in which the sites of the lattice are specified by the dual (reciprocal) lattice vector
 ${\bf q}\cdot {\bf p}=2\pi m$, where $m= \pm 1, \pm 2, \pm 3 \cdots$.
The dual fundamental cell $F^*$ has volume $v_{F^*}=(2\pi)^d/v_F$.
This implies that the number density $\rho$ of $\Lambda$
is related to the number  density $\rho_*$ of the dual lattice $\Lambda^*$
via the expression $\rho \rho_*=1/(2\pi)^d$.
A {\it periodic} point process is a
more general notion than a lattice because it is
is obtained by placing a fixed configuration of $N$ points (where $N\ge 1$)
within one fundamental cell of a lattice $\Lambda$, which 
is then periodically replicated. Thus, the point process is still
periodic under translations by $\Lambda$, but the $N$ points can occur
anywhere in the chosen fundamental cell.

Common $d$-dimensional lattices include the {\it hypercubic} $\mathbb{Z}^d$, 
{\it checkerboard} $D_d$ and 
{\it root} $A_d$ lattices,
defined, respectively, by
\begin{equation}
\mathbb{Z}^d=\{(x_1,\ldots,x_d): x_i \in {\mathbb{ Z}}\} \quad \mbox{for}\; d\ge 1
\end{equation}
\begin{equation}
D_d=\{(x_1,\ldots,x_d)\in \mathbb{Z}^d: x_1+ \cdots +x_d ~~\mbox{even}\} \quad \mbox{for}\; d\ge 3
\end{equation}
\begin{equation}
A_d=\{(x_0,x_1,\ldots,x_d)\in \mathbb{Z}^{d+1}: x_0+ x_1+ \cdots +x_d =0\} \quad \mbox{for}\; d\ge 1
\end{equation}
where $\mathbb{Z}$ is the set of integers ($\ldots -3,-2,-1,0,1,2,3\ldots$)
and $x_1,\ldots,x_d$ denote the components of a lattice vector
of either $\mathbb{Z}^d$ or $D_d$ and $x_0,x_1,\ldots,x_d$ denote
a lattice vector of $A_d$. The $d$-dimensional lattices $\mathbb{Z}^d_*$, $D_d^*$ and $A_d^*$ are
the corresponding dual lattices; see Ref.~\cite{Co93} for definitions.
The dual lattice $\mathbb{Z}^d_*$ is also a hypercubic lattice (even if the lattice spacing
is $2 \pi$ times the lattice spacing of $\mathbb{Z}^d$) and hence we say that the hypercubic
lattice $\mathbb{Z}^d$ is equivalent (similar) to its dual lattice $ \mathbb{Z}^d_*$, 
i.e., $\mathbb{Z}^d\equiv \mathbb{Z}^d_*$.
Following Conway and Sloane \cite{Co93}, we say that
two lattice are {\it equivalent} or {\it similar} if one becomes identical
to the other by possibly a rotation, reflection and change of scale,
for which we use the symbol $\equiv$. In fact, the hypercubic
lattice $\mathbb{Z}^d$ is characterized by the stronger property
of {\it self-duality}. A {\it self-dual} lattice is one with an {\it identical} dual lattice at
density $\rho = \rho_* = 1/(2\pi)^{d/2}$, i.e., without any rotation,
reflection, or change of scale \cite{duality}.
The $A_d$ and $D_d$ lattices can be regarded to be $d$-dimensional generalizations
of the face-centered-cubic (fcc) lattice because this three-dimensional
lattice is defined by $A_3 \equiv D_3$. 
In one dimension, $A_1=A_1^*$ (equality meaning self-duality) are identical to the integer lattice $\mathbb{Z}^1 =\mathbb{Z}$.
In two dimensions, $A_2 \equiv A_2^*$ defines the triangular lattice.
In three dimensions, $A_3^* \equiv D_3^*$ defines the body-centered-cubic (bcc)
lattice. The $d$-dimensional laminated lattice $\Lambda_d$ \cite{Co93,laminate} is of special
interest. In dimensions 8 and 24,  $\Lambda_8\equiv E_8$, where $E_8$ is 
the {\it self-dual} root lattice, and $\Lambda_{24}$ is the self-dual
Leech lattice,
are remarkably symmetric and believed to be the densest sphere
packings in those dimensions \cite{Co09}. 
 Thus, $E_8=E_8^*$
and $\Lambda_{24}=\Lambda_{24}^*$.
The laminated lattice $\Lambda_{16}$,
called the Barnes-Wall lattice, and the Coxeter-Todd lattice $K_{12}$ are thought to be the
densest lattice packings in sixteen and twelve dimensions, respectively.

Note that for a single periodic point configuration at number density $\rho$, the radial pair correlation function can be written as 
\begin{equation}
g_2(r)=  \sum_{i=1}^{\infty}\frac{Z_i}{\rho s_1(r_i)}\;\delta(r-r_i),
\label{period}
\end{equation}
where $Z_i$ is the coordination number at radial distance $r_i$
(number of points that are exactly at a distance $r=r_i$ from
a point of the point process) 
such that $r_{i+1} > r_i$ and $\delta(r)$ is a radial Dirac delta
function. For cases in which each point is equivalent to any other, 
which includes all lattices and some periodic point processes,
the coordination numbers $Z_i$ are integers. For point processes
for which the point are generally inequivalent, $Z_i$ should
be interpreted as the {\it expected} coordination number
and hence will generally be a non-integer. Substitution of (\ref{period}) into (\ref{cum})
gives the coordination function for such a periodic configuration as
\begin{equation}
Z(R)= \sum_{i=1}^M Z_i,
\end{equation}
where $M$ is the smallest integer for which $r_{M+1} >R$.

Consider any discrete set of points with position vectors
$Y=\{{\bf r}_1, {\bf r}_2,\ldots\}$ in $\mathbb{R}^d$. Associated with each point ${\bf r}_i \in Y$
is its {\em Voronoi cell}, ${\cal V}({\bf r}_i)$,
which is defined to be the region of space nearer
to the point at ${\bf r}_i$ than to any other point ${\bf r}_j$ in the set, i.e.,
\begin{equation}
{ \cal V}({\bf r}_i)=\{{\bf x}: |{\bf x}-{\bf r}_i| \le |{\bf x}-{\bf r}_j|\;
\mbox{for all}\; {\bf r}_j \in Y\}.
\end{equation}
The Voronoi cells are convex polyhedra whose interiors
are disjoint, but share common
faces, and therefore the union of all of the polyhedra
is the whole of $\mathbb{R}^d$. This partition of space is called the {\em Voronoi} tessellation.
 The vertices of the Voronoi polyhedra are the points whose distance from the points
$Y$ is a local maximum.
While the Voronoi polyhedra of a lattice are congruent to one another,
the Voronoi polyhedra of a non-Bravais lattice are not identical
to one another. A hole in a lattice is a point in $\mathbb{R}^d$  whose distance to the nearest lattice point is a local maximum. A {\it deep} hole is one whose distance to a lattice point is a global maximum. 
The distance ${\cal R}_c$ to the deepest hole of a lattice is the {\it covering radius}
and is equal to the {\it circumradius} of the associated Voronoi cell
(the radius of the smallest circumscribed sphere). 

\begin{figure}
\centerline{\includegraphics[width=1.8in,,keepaspectratio,clip=]{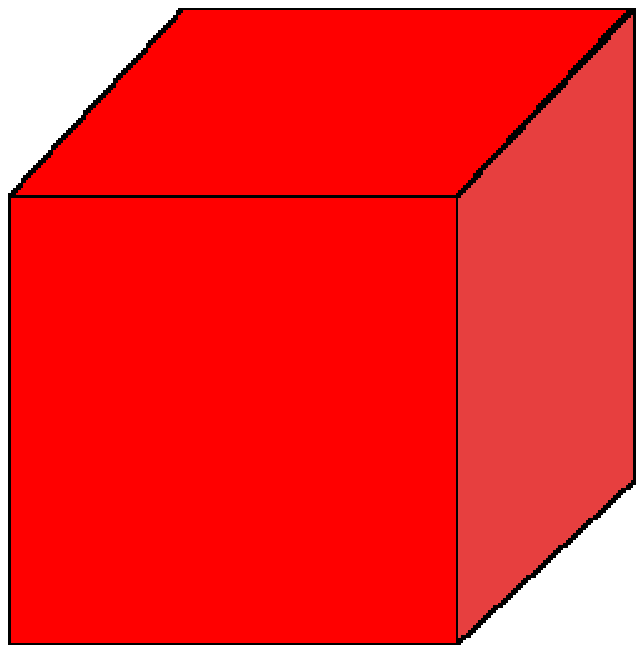}
\includegraphics[width=1.6in,keepaspectratio,clip=]{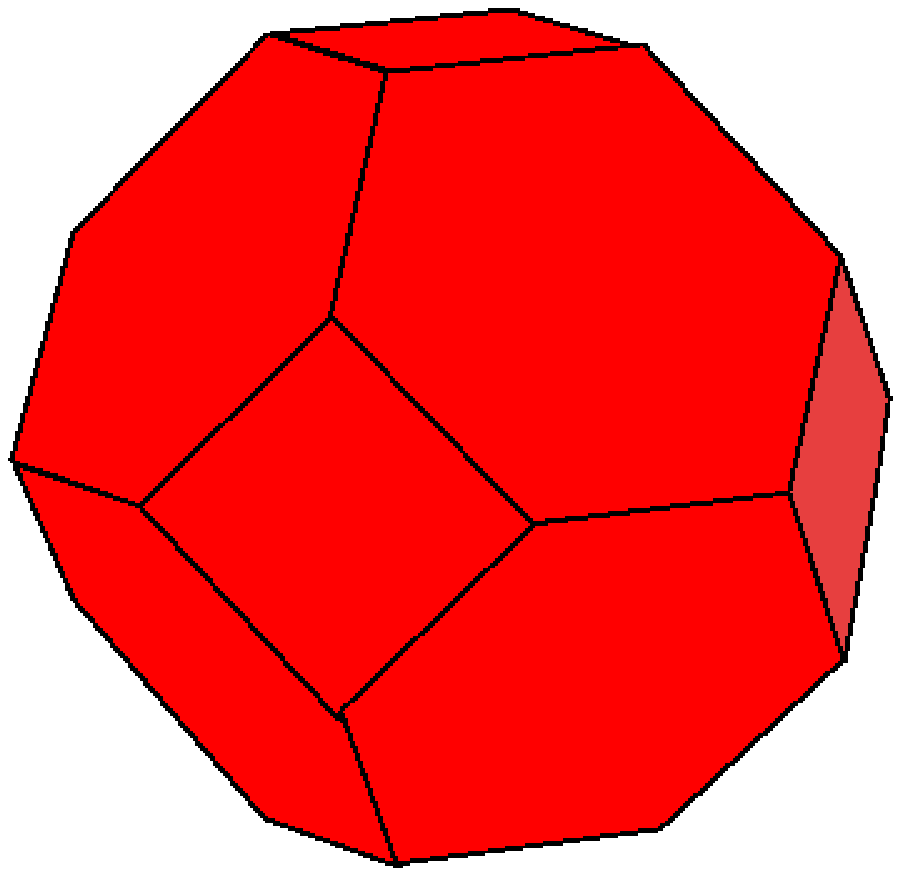}
\includegraphics[width=1.8in,keepaspectratio,clip=]{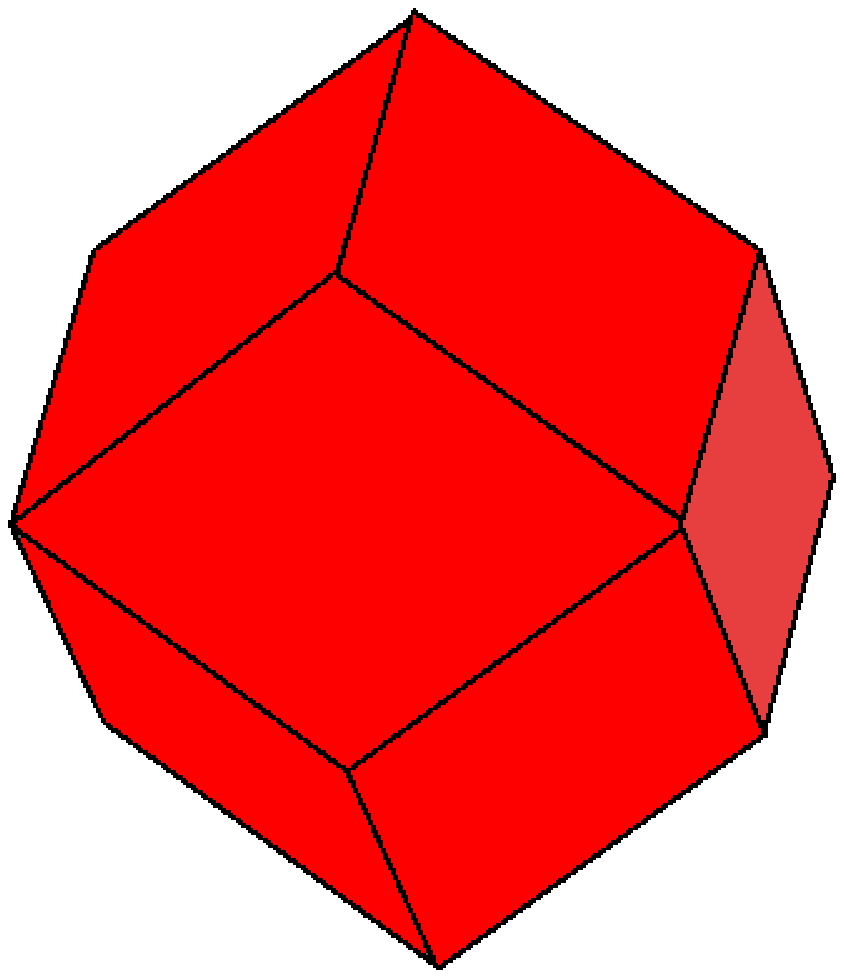}}
\caption{(Color online) Voronoi cells in $\mathbb{R}^3$ for simple cubic, body-centered cubic, and face-centered cubic lattices
are the cube (left), truncated octahedron (middle), and rhombic dodecahedron (right),
as adapted from Ref. \cite{To02a}.
The truncated octahedron is composed of  six square and eight regular hexagonal faces.
The rhombic dodecahedron is composed of twelve rhombus-shaped faces.}
\label{reg3d-vor}
\end{figure}

In the case of the $d$-dimensional simple cubic (hypercubic) lattice $\mathbb{Z}^d$, the Voronoi cell
is a hypercube and there is only one type
of hole with covering radius ${\cal R}_c= \sqrt{d}/2$, assuming unit number density $\rho=1$.
Figure \ref{reg3d-vor} shows the Voronoi cell in the case $d=3$ as
well as the corresponding Voronoi cells for the three-dimensional
body-centered cubic and face-centered cubic lattices. Note that
the truncated octahedron is the most spherically symmetric
of the three Voronoi cells shown in Fig.~\ref{reg3d-vor} \cite{sym}.

A sphere packing $P$ in $d$-dimensional Euclidean space $\mathbb{R}^d$ is a collection of  $d$-dimensional nonoverlapping congruent spheres.
The {\it packing density} or, simply, density $\phi(P)$ of a sphere packing is the fraction of
space $\mathbb{R}^d$ covered by the spheres. For spheres of diameter $D$
and number  density $\rho$, the density is given by
\begin{equation}
\phi=\rho v_1(D/2),
\end{equation}
where
\begin{equation}
v_1(R) = \frac{\pi^{d/2}}{\Gamma(1+d/2)} R^d
\label{v1}
\end{equation}
is the volume of a $d$-dimensional sphere of radius $R$. 

A packing is {\it saturated} if there is no space available to add another sphere
 without overlapping the existing particles. We denote the packing density
of such a packing by $\phi_s$.
A {\it lattice packing} $P_L$ is one in which  the centers of nonoverlapping spheres
are located at the points of $\Lambda$.
Thus, the density of a lattice packing $\phi^L$ consisting
of spheres of diameter $D$ is given by
\begin{equation}
\phi^L= \frac{v_1(D/2)}{v_F},
\end{equation}
where  $v_F$ is the volume of a fundamental cell.
A {\it periodic} packing of congruent spheres 
is obtained by placing a fixed configuration of $N$ sphere centers (where $N\ge 1$)
within one fundamental cell of a lattice $\Lambda$, which is then periodically replicated without overlaps. The packing density of a  periodic packing is given by
\begin{equation}
\phi=\frac{N v_1(D/2)}{v_F}=\rho v_1(D/2),
\label{den}
\end{equation}
where $\rho=N/v_F$ is the number density.

\section{Problem Statements and Background}
\label{problems}

\subsection{Sphere Packing Problem}
\label{sphere}

The sphere packing problem seeks to answer the following
question: Among all packings of congruent spheres,
what is the maximal packing density $\phi_{\mbox{\scriptsize max}}$, i.e., largest
fraction of $\mathbb{R}^d$ covered by the spheres,
and what are the corresponding arrangements of the spheres \cite{Ro64,Co93}?
More precisely, the maximal density is defined by
\begin{equation}
\phi_{\mbox{\scriptsize max}}= \sup_{P\subset \mathbb{R}^d} \phi(P),
\end{equation}
where the supremum is taken over all packings in $\mathbb{R}^d$.
The sphere packing problem is of great fundamental and practical interest,
and arises in a variety of contexts, including classical ground states
of  matter in low dimensions \cite{Chaik95},
the famous Kepler conjecture for $d=3$ \cite{Ha05}, error-correcting
codes \cite{Sh48,Co93} and spherical codes \cite{Co93}.

The optimal solutions are known only for the first three space
dimensions \cite{Ha05}. For $4 \le  d \le 9$,
the densest known packings are Bravais lattice
packings  \cite{Co93}. For example, the ``checkerboard" lattice $D_d$, which is
the $d$-dimensional generalization of the fcc lattice
(densest packing in $\mathbb{R}^3$), is believed
to be optimal in $\mathbb{R}^4$ and $\mathbb{R}^5$.
The remarkably symmetric self-dual $E_8$ and Leech lattices in $\mathbb{R}^8$
and $\mathbb{R}^{24}$, respectively, are most
likely the densest packings in these dimensions \cite{Co09}.
Table~\ref{pack} summarizes the densest known packings in selected
dimensions.

\begin{table}[bthp]
\caption{Best known solutions to the sphere packing problem
in selected dimensions; see Conway and Sloane \cite{Co93} for details. }
\label{pack}
\renewcommand{\baselinestretch}{1.2} \small \normalsize
\centering
\begin{tabular} {|c|c|c|}
\multicolumn{3}{c}{~} \\\hline
 {Dimension, $d$} & Packing & Packing density, $\phi$\\ \hline
1 &  $A_1^*= {\mathbb{Z}}$ & 1 \\

2 & $A_2^*\equiv A_2$  & $\pi/\sqrt{12}=0.906899\ldots$\\

3 &  $A_3\equiv D_3$ &  $\pi/\sqrt{18}=0.740480\ldots$ \\

4 &  $D_4\equiv D_4^*$ & $\pi^2/16=0.616850\ldots$ \\

5 &  $D_5$ &   $2\pi^2/(30\sqrt{2})=0.465257\ldots$\\

6 &  $E_6$ & $3\pi^3/(144\sqrt{3})=0.372947\ldots$\\ 
7 &  $E_7$ & $\pi^3/105= 0.295297\ldots$\\ 
8 &  $E_8= E_8^*$ & $\pi^4/384=0.253669\ldots$\\ 
9 &   $\Lambda_9$ &  $\sqrt{2}\pi^4/945= 0.145774\ldots$\\ 
10&   $P_{10c}$ & $\pi^5/3072=0.099615\ldots$\\ 
12&  $\Lambda^{max}_{12}$ &$\pi^6/23040=0.041726\ldots$\\ 
16&   $\Lambda_{16}$ &$\pi^8/645120=0.014708\ldots$\\ 
24&  $\Lambda_{24}=\Lambda_{24}^*$ &$\pi^{24}/479001600=0.001929\ldots$  
\\ \hline
\end{tabular}
\end{table}

For large $d$, the best that one can do theoretically
is to devise  upper and lower bounds on
$\phi_{\mbox{\scriptsize max}}$ \cite{Co93,Co03,To06b}.
For example, Minkowski \cite{Mi05} proved that the maximal density
$\phi^L_{\mbox{\scriptsize max}}$ among all Bravais lattice packings 
for $d \ge 2$ satisfies the lower bound 
\begin{equation}
\phi^L_{\mbox{\scriptsize max}} \ge \frac{\zeta(d)}{2^{d-1}},
\label{mink}
\end{equation}
where $\zeta(d)=\sum_{k=1}^\infty k^{-d}$ is the Riemann zeta function.
It is seen that for large values of $d$,
the asymptotic behavior of the {\it nonconstructive} Minkowski lower bound is controlled by $2^{-d}$.
Note that the density of a {\it saturated} packing of congruent spheres
in $\mathbb{R}^d$ for all $d$ satisfies 
\begin{equation}
\phi \ge \frac{1}{2^d},
\label{sat}
\end{equation}
which has the same dominant exponential term as (\ref{mink}).
This is a rather weak lower bound on the density of saturated packings
because there exists a  disordered but {\it unsaturated packing construction}
in $\mathbb{R}^d$, known as the ``ghost"
RSA packing \cite{To06a}, that achieves the density $2^{-d}$ in any dimension.
We will employ these results in Sec.~\ref{improve}.
It is also known that there are saturated packings
in $\mathbb{R}^d$ with densities that  exceed the scaling $2^{-d}$ \cite{To06d}, as we will
discuss in Sec.~\ref{cov}.
In the large-dimensional limit, Kabatiansky and Levenshtein   \cite{Ka78} showed that the maximal density
is bounded from above according to the asymptotic upper bound
\begin{equation}
\phi_{\mbox{\scriptsize max}} \le \frac{1}{2^{0.5990\,d}}.
\label{kab}
\end{equation}

We will employ sphere-packing solutions to obtain
heretofore unattained results for both the covering
and quantizer problems. In particular, we obtain coverings
and quantizers utilizing disordered saturated packings 
in Secs.~\ref{results-cover} and \ref{sat-quant}, respectively.
In Sec.~\ref{results-quant}, we use the densest lattice packings
to derive improved upper bounds on  the quantizer error.

\subsection{Number Variance Problem, Hyperuniformity and Epstein Zeta Function}
\label{num-var}

We denote by $\sigma^2(A)$
the variance in the number of points $N(A)$ contained within a window
$A \subset \mathbb{R}^d$. The number variance $\sigma^2(A)$ for a specific choice
of $A$ is necessarily a positive number and is
generally related to the {\it total  
correlation function} $h({\bf r})=g_2({\bf r})-1$ for
a translationally invariant point process \cite{To03a}, where $g_2$ is
the pair correlation function defined in Sec.~\ref{def}.
In the special case of a spherical window of radius $R$ in $\mathbb{R}^d$,
the number variance is explicitly given by
\begin{eqnarray}
\sigma^2(R)=\rho v_1(R) \Bigg[ 1+\rho \int_{\mathbb{R}^d} h({\bf r}) \alpha(r; R) \, d{\bf r}\Bigg],
\label{variance}
\end{eqnarray}
$\alpha(r;R)$ is the dimensionless volume common to two spherical windows of radius $R$
(in units of the volume of a spherical window of radius $R$, $v_1(R)$)
whose centers are separated by a distance $r$.
We will call $\alpha(r;R)$ the {\it scaled intersection volume}, which will play an important
role in this paper. The scaled intersection volume  has the support
$[0,2R]$, the range $[0,1]$, and the following alternative integral representation \cite{To06b}:
\begin{eqnarray}
\alpha(r;R) = c(d) \int_0^{\cos^{-1}[r/(2R)]} \sin^d(\theta) \, d\theta,
\label{alpha}
\end{eqnarray}
where $c(d)$ is the $d$-dimensional constant given by
\begin{eqnarray}
c(d)= \frac{2 \Gamma(1+d/2)}{\pi^{1/2} \Gamma[(d+1)/2]}.
\label{C}
\end{eqnarray}

\begin{figure}
\centerline{ \includegraphics[width=3.5in]{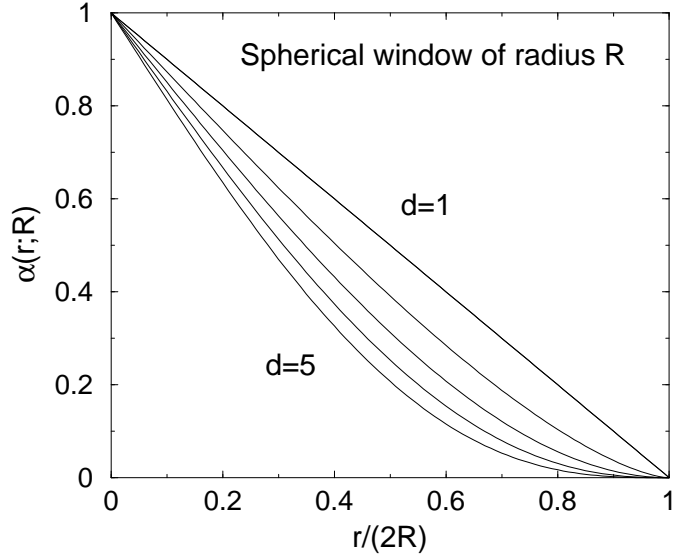}}
\caption{ The scaled intersection
volume $\alpha(r;R)$ for spherical windows of radius $R$
as a function of $r$ for the first
five space dimensions. The uppermost curve is for $d=1$
and lowermost curve is for $d=5$. }
\label{intersection}
\end{figure}

Torquato and Stillinger \cite{To06b} found the following  series representation
of the scaled intersection volume $\alpha(r;R)$ for $r \le 2R$
and for any $d$:
\begin{eqnarray}
\alpha(r;R)=1- c(d) x+ 
c(d) \sum_{n=2}^{\infty}
(-1)^n \frac{(d-1)(d-3) \cdots (d-2n+3)}
{(2n-1)[2 \cdot 4 \cdot 6 \cdots (2n-2)]} x^{2n-1},
\label{series}
\end{eqnarray}
where $x=r/(2R)$. For even dimensions, relation (\ref{series}) is
an infinite series because it involves transcendental functions, but for odd dimensions, the series truncates such
that $\alpha(r;R)$ is a univariate polynomial of degree $d$. 
For example, in two and three dimensions, respectively, the scaled intersection
volumes are given by
\begin{equation} 
\alpha(r;R) =   \frac{2}{\pi} \left[ \cos^{-1}\left(\frac{r}{2R}\right) - \frac{r}{2R}
\left(1 - \frac{r^2}{4R^2}\right)^{1/2} \right]\Theta(2R-r) \quad (d=2),
\end{equation}
\begin{equation}
\alpha(r;R)= \left[ 1 -\frac{3}{4}\frac{r}{R}+\frac{1}{16} \left(\frac{r}{R}\right)^3 \right] \Theta(2R-r) \quad (d=3),
\end{equation}
where
\begin{equation}
\Theta(x) =\Bigg\{{0, \quad x<0,\atop{1, \quad x \ge 0,}}
\label{heaviside}
\end{equation}
is the Heaviside step function.
Figure \ref{intersection} provides plots of $\alpha(r;R)$ as a function of $r$ for the first five space dimensions. For any dimension, $\alpha(r;R)$
is a monotonically decreasing function of $r$. At a fixed
value of $r$ in the interval $(0,2R)$, $\alpha(r;R)$
is a monotonically decreasing function of the dimension $d$.

For large $R$, it has been proved that $\sigma^2(R)$ cannot grow more slowly than $\gamma R^{d-1}$, where $\gamma$ is a positive constant \cite{Beck87}. We note that point processes (translationally invariant or not)
for which $\sigma^2(R)$ grows more slowly than the window volume (i.e.,
as  $R^{d}$) for large $R$ are examples of {\it hyperuniform} (or superhomogeneous) point patterns 
\cite{To03a}.  For hyperuniform point processes
in which the number variance grows like the surface area of the window, one has
\begin{equation}
\sigma^2(R)=2^d \eta
\left[ B \left(\frac{R}{{\cal D}}\right)^{d-1} +{\cal O} \left(\frac{R}{{\cal D}}\right)^{d-2}\right] \qquad
R \rightarrow \infty,
\end{equation}
where 
\begin{equation}
B= \frac{\eta\; c(d)}{2{\cal D} v_1({\cal D}/2)} \int_{\mathbb{R}^d} h({\bf r}) r d{\bf r},
\label{b}
\end{equation}
is a dimensionless constant  with $c(d)$ defined by (\ref{C}),
\begin{equation}
\eta=\rho v_1({\cal D}/2)
\end{equation}
is a dimensionless density, and ${\cal D}$ represents some ``microscopic" length scale, such
as the minimum pair separation distance in  a packing or the mean nearest-neighbor distance.
This class of hyperuniform point processes includes all periodic point patterns, quasicrystals that 
possess Bragg peaks, and disordered hyperuniform point patterns in which the
pair correlation functions decay exponentially fast to unity \cite{To03a}.

It has been shown that finding the point process that minimizes the number variance
$\sigma^2(R)$ is equivalent to finding the ground state of a certain repulsive pair potential
with compact support \cite{To03a}. 
Specifically, by invoking a volume-average interpretation of the 
number variance problem valid
for a single realization of a point process, Torquato and Stillinger found \cite{To03a}:
\begin{eqnarray}\label{volavginterp}
\sigma^2(R) = 2^d\eta B_N(R)\left(\frac{R}{{\cal D}}\right)^{d-1},
\end{eqnarray}
where
\begin{equation}
B_N(R) = \frac{R}{{\cal D}}\left[1-2^d\eta \left(\frac{R}{{\cal D}}\right)^d + 
\frac{1}{N}\sum_{i\neq j }^N \alpha(r_{ij}; R)\right].
\label{Bn}
\end{equation}
The asymptotic coefficient $B$ defined by (\ref{b}) for a hyperuniform point pattern is then 
related to $B_N(R)$ by the expression
\begin{equation}
B = \lim_{L\rightarrow +\infty} \frac{1}{L}\int_0^L B_N(R) dR.
\end{equation}
These results imply that the asymptotic coefficient $B$ obtained in (\ref{b}) 
involves an average
over small-scale fluctuations in the number variance with length scale on the order of the
mean separation between points \cite{To03a}.
In the special case of a (Bravais) lattice $\Lambda$,
one can express the rescaled surface-area coefficient as follows:
\begin{eqnarray}
\eta^{1/d} B = \frac{\pi^{(d-1)/2} 2^{d-1} \left[\Gamma(1+d/2)\right]^{1-1/d}}{v_F^{1+1/d}} \sum_{\mathbf{q}\neq\mathbf{0}} \frac{1}{\Vert\mathbf{q}\Vert^{d+1}},
\label{num}
\end{eqnarray}
where we recall that $v_F$ is the volume of the fundamental cell of the lattice $\Lambda$
and ${\bf q}$ represents a lattice vector
in the dual (or reciprocal) lattice $\Lambda^*$. 
The rescaled coefficient $\eta^{1/d}B$ renders the result
independent of the length scale in the lattice \cite{To03a}.

Finding the lattice
that minimizes $B$  is directly related 
to an outstanding problem in number theory, namely, finding
the minima of the Epstein zeta function $Z_{\Lambda}(s)$ \cite{Sa06} defined by
\begin{equation}
Z_{\Lambda}(s)=\sum_{\bf p \neq {\bf 0}} \frac{1}{\Vert{\bf p}\Vert^{2s}},
\label{epstein}
\end{equation}
where ${\bf p}$ is a lattice vector of the lattice $\Lambda$.
Note  that the {\it dual} of the lattice
that minimizes the Epstein zeta function at $s=(d+1)/2$ among all lattices will minimize the scaled asymptotic number-variance coefficient (\ref{num}) among lattices \cite{To03a,Za09}. 
Certain duality relations have been derived  that establish rigorous
upper bounds on the energies of such ground states and help to identify energy-minimizing lattices \cite{To08a}.
Because $Z_{\Lambda}(s)$ is globally minimized for $d=1$ by the
integer lattice \cite{To03a} and is minimized 
for $d=2$ among all
lattices by the triangular lattice \cite{Ra53},
it has been conjectured that the Epstein zeta function for $ s >0$ is minimized among lattices
by the maximally dense lattice packing \cite{En64,Chiu97}.
Sarnak and Str{\" o}mbergsson \cite{Sa06} have proved that the conjecture
cannot be generally true, but for $d=4$, $8$ and $24$, the densest
lattice packing is a strict local minimum. Since as $s \rightarrow +\infty$,
the minimizer of Epstein zeta function is the densest sphere packing in $\mathbb{R}^d$ for any $d$, it is likely that in the high-dimensional limit the minimizers of this
function are non-lattices, namely, disordered sphere packings \cite{To06b}.

\begin{table}
\caption{Best known solutions to the
asymptotic number variance  problem in selected dimensions. 
Values reported for $d=1,2$ and 3 and $d=4$-8 are taken from Torquato and Stillinger \cite{To03a}
and Zachary and Torquato \cite{Za09}, respectively. Values
reported for $d=12,16$ and 24 have been determined in the present work. }
\label{var}
\renewcommand{\baselinestretch}{1.2} \small \normalsize
\centering
\begin{tabular} {|c|c|c|}
\multicolumn{3}{c}{~} \\\hline
 {Dimension, $d$} & Structure& Scaled Coefficient, $\eta^{1/d}B$ \\ \hline
1 &  $A_1^*=\mathbb{Z}$ & 0.083333 \\

2 & $A_2^*\equiv A_2$  & 0.12709 \\

3 &  $A_3^*\equiv D_3^*$ & 0.15560 \\

4  & $D_4^*\equiv D_4$  & 0.17488 \\
5  & $\Lambda_5^{2*}$ & 0.19069 \\
6  & $E_6^*$  & 0.20221 \\
7  & $D_7^+$ & 0.21037 \\
8  & $E_8^*= E_8 = D_8^+$  & 0.21746 \\ 
12  & $K_{12}^*\equiv K_{12}$ &  0.24344 \\
16 & $\Lambda_{16}^*\equiv \Lambda_{16}$  & 0.25629 \\
24 & $\Lambda_{24}^*=\Lambda_{24}$  &  0.26775 \\ \hline
\end{tabular}
\end{table}

In Table~\ref{var}, we tabulate the best  known solutions to the
asymptotic number variance  problem in selected dimensions.
Values for the first three space dimensions were given
in Ref. \cite{To03a} and those for $d=4-8$ were provided
in Ref. \cite{Za09}. The values reported for $d=12, 16$
and 24 were ascertained using efficient algorithms
based on alternative number-theoretic
representations of the Epstein zeta function
$Z_{\Lambda}(s)$ \cite{Sa06}  for the corresponding densest known lattice
packings for $s=(d+1)/2$ and then using the duality
relations connecting it to the asymptotic
surface-area coefficient for the number variance.
Appendix \ref{num-Ep} provides details for
these computations.

\subsection{Covering Problem}
\label{cov}

\begin{figure}
 \centerline{ \includegraphics[width=2.7in,,keepaspectratio,clip=,angle=90]{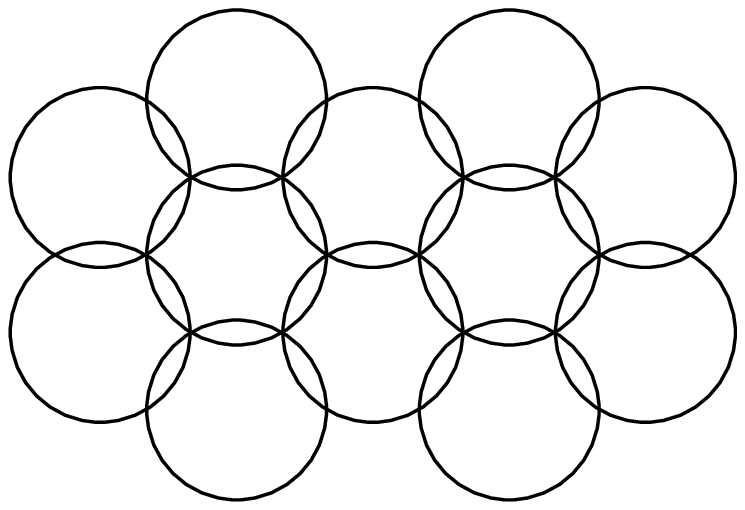}
\includegraphics[width=2.2in,,keepaspectratio,clip=]{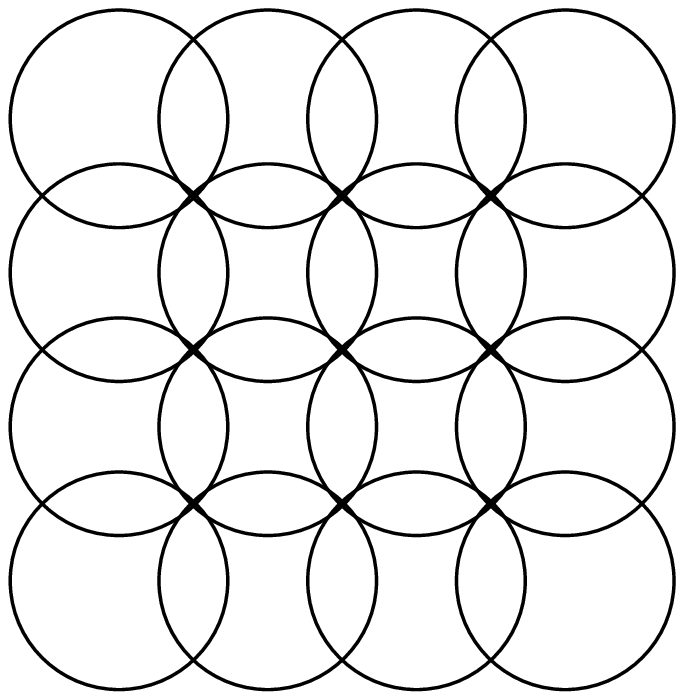}}
\caption{ Coverings of the plane with overlapping circles centered
on the triangular lattice (left panel) and the square lattice (right panel). 
The triangular lattice $A_2\equiv A_2^*$ provides the best covering among all point processes
at unit number density $\rho=1$ with $\theta=2\pi/(3\sqrt{3})=1.2092\ldots$.
This is to be compared to the square lattice $\mathbb{Z}^2$ with $\theta=\pi/2=1.5708\ldots$.}
\label{sq-tri}
\end{figure}

Surround each of the points of a point process $\cal P$ in $\mathbb{R}^d$ 
by congruent overlapping spheres
of radius $R$ such that the spheres cover the space.
The {\it covering density} $\theta$ is defined as follows:
\begin{equation}
\theta=\rho v_1(R),
\label{covering}
\end{equation}
where $v_1(R)$ is given by (\ref{v1}).
The {\it covering problem} asks for the arrangement of points
with the least density $\theta$. We define 
the covering radius ${\cal R}_c$ for any configuration 
of points in $\mathbb{R}^d$ to be the minimal radius of the overlapping
spheres to cover $\mathbb{R}^d$. Figure ~\ref{sq-tri} shows
two examples of coverings in the plane.

The covering density associated with $A_d^*$ at unit number density $\rho=1$ 
is known exactly for any dimension $d$ \cite{Co93}:
\begin{equation}
\theta =v_1(1) \sqrt{d+1}\left[\frac{d(d+2)}{12(d+1)}\right]^{d/2}.
\end{equation}
For the hypercubic lattice $\mathbb{Z}^d$ at $\rho=1$,
\begin{equation}
\theta =v_1(1) \frac{d^{d/2}}{2^d}.
\end{equation}
Thus the ratio of the covering density for $A_d^*$
to that of $\mathbb{Z}^d$ is given by
\begin{equation}
\frac{\theta(A_d^*)}{\theta(Z^d)} = \frac{\sqrt{d+1}}{3^{d/2}}\left[\frac{d+2}{d+1}\right]^{d/2}.
\end{equation}
For large $d$, this ratio becomes
\begin{equation}
\frac{\theta(A_d^*)}{\theta(Z^d)} \sim \frac{\sqrt{d e}}{3^{d/2}},
\end{equation}
and thus we see that $A_d^*$ provides exponentially
thinner coverings than that of $\mathbb{Z}^d$ in the large-$d$ limit.
We note that in this asymptotic limit,
\begin{equation}
\theta(A_d^*) \sim \left(\frac{e}{\pi}\right)^{1/2}  
\left(\frac{e\pi}{6}\right)^{1/2}= 0.8652559792\ldots (1.193016780\ldots)^d.
\label{A_n^*}
\end{equation}
Until recently, $A_d^*$ was the best known lattice covering in all dimensions
$d \le 23$. However, for most dimensions in the
range $ 6 \le \theta \le 15$, Sch{\"u}rmann and Vallentin \cite{Sc06}
have discovered other lattice coverings that are slightly thinner than those
for $A_d^*$. Table \ref{coverings}  provides the best known solutions to the covering problem
in selected dimensions.

\begin{table}[bthp]
\caption{Best known solutions to the 
covering problem in selected dimensions. Values
reported are taken from Conway and Sloane \cite{Co93}, except
for $d=7,8$ and 9, which were obtained from  Sch{\"u}rmann and Vallentin \cite{Sc06}. It is only in one and two dimensions that these
solutions have been proved to be globally optimal \cite{Co93}.}
\label{coverings}
\renewcommand{\baselinestretch}{1.2} \small \normalsize
\centering
\begin{tabular} {|c|c|c|}
\multicolumn{3}{c}{~} \\\hline
 {Dimension, $d$} & Covering & Covering Density, $\theta$ \\ \hline
1 & $A_1^*= \mathbb{Z}$     & 1 \\
2 & $A_2^*\equiv A_2$   & 1.2092 \\
3 & $A_3^*\equiv D_3^*$ & 1.4635 \\
4 & $A_4^*$             & 1.7655 \\
5 & $A_5^*$             & 2.1243 \\
6 & $L_6^{c1}$          & 2.4648\\ 
7 & $L_7^c$             &  2.9000\\ 
8 & $L_8^c$             & 3.1422 \\ 
9 & $A^5_9$             & 4.3401\\ 
10& $A^*_{10}$          & 5.2517 \\ 
12& $A^*_{12}$          & 7.5101\\ 
16& $A^*_{16}$          & 15.3109 \\ 
17& $A^*_{17}$          & 18.2878\\
18& $A^*_{18}$          &21.8409\\
24& $\Lambda_{24}=\Lambda_{24}^*$      & 7.9035
\\ \hline
\end{tabular}
\end{table}

\begin{figure}
 \centerline{ \includegraphics[width=2.5in,,keepaspectratio,]{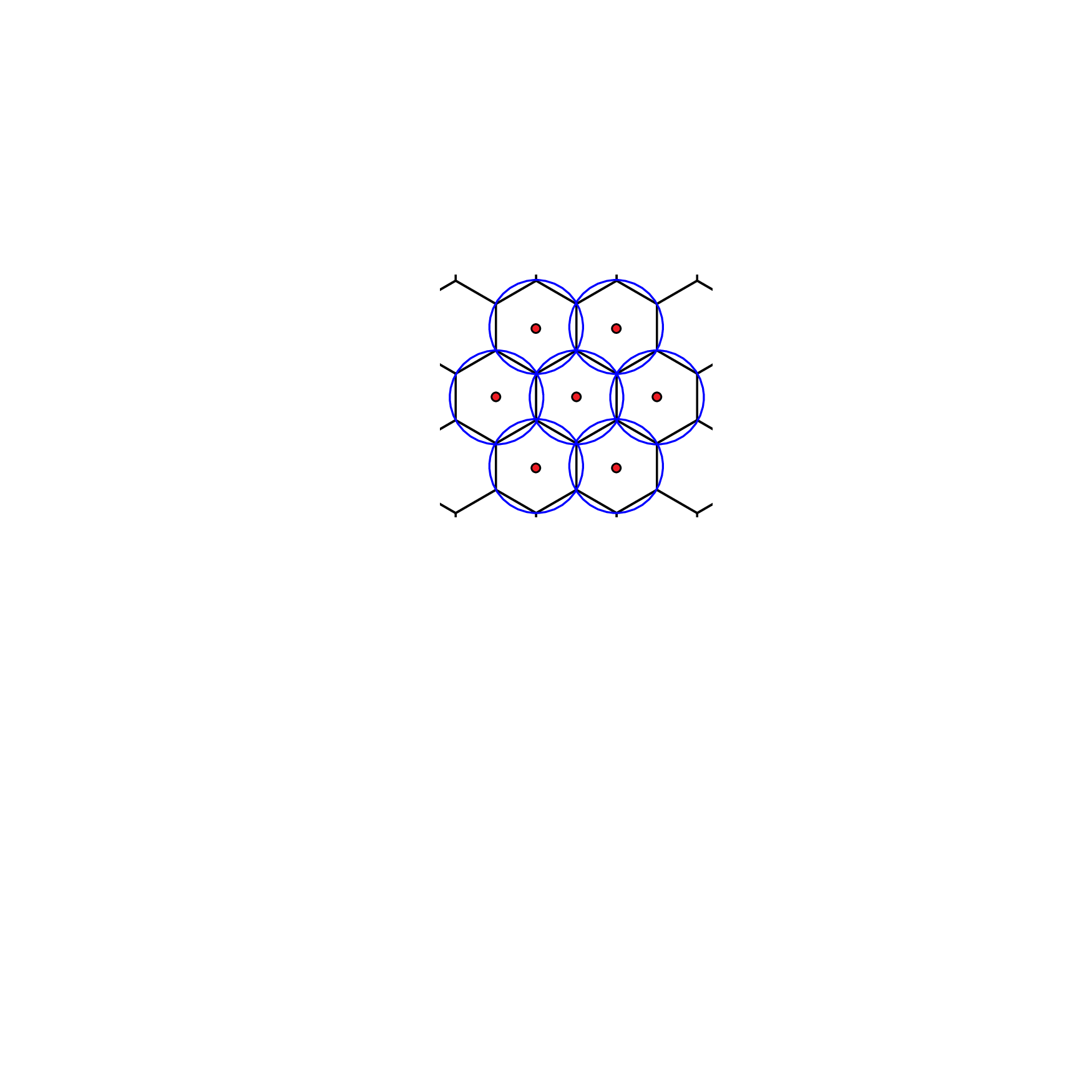}
\hspace{0.5in}\includegraphics[width=2.2in,,keepaspectratio,clip=]{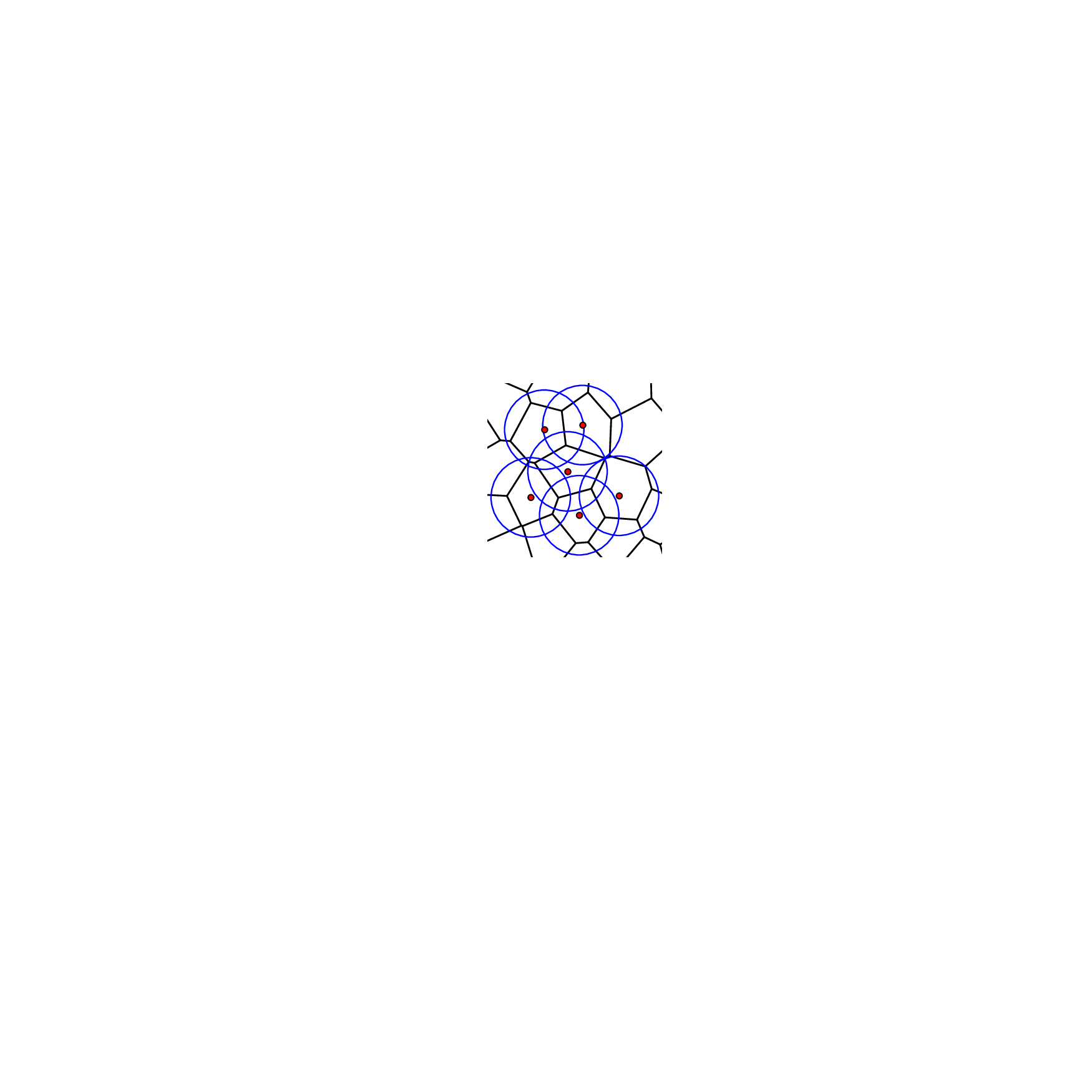}}
\caption{(Color online) For lattices or periodic point process in which each
point is equivalent, the Voronoi cells are congruent to one another,
the Voronoi centroids coincide with the points of the point process,
and the covering radius ${\cal R}_c$ is equal to the circumradius of the associated Voronoi
cell. For an irregular point pattern, generally, the Voronoi cells are not congruent
to one another, the Voronoi centroids do not coincide with the points of the point process, and
 and the covering radius ${\cal R}_c$ 
is not equal to the circumradius of the associated Voronoi
cell. These two instances are illustrated in two dimensions for the triangular
lattice and an irregular point pattern.}
\label{tri-RSA-cov}
\end{figure}

Until the present work, there were  no known explicit non-lattice constructions 
possessing  covering densities smaller than those of the best lattice coverings 
in any dimension $d$ \cite{Co93}.
In Sec.~\ref{results-quant}, we provide evidence that certain disordered
point patterns give thinner coverings than the best known
lattice coverings beginning at $d=17$. There is a fundamental
difference between coverings associated with point patterns
that have identical Voronoi cells (i.e., lattices and
periodic point patterns in which each point is equivalent)
and those point processes whose Voronoi cells are generally
different (e.g., irregular point processes).
This salient point is illustrated in Fig.~\ref{tri-RSA-cov}
and explained in the corresponding caption.

Rogers showed that (possibly nonlattice) coverings
exist with
\begin{equation}
\theta \le 5d+ d \ln(d)+d \ln(\ln(d)).
\label{cover-up}
\end{equation}
for $d \ge 2$. This is a nonconstructive upper bound.
This upper bound provides a substantially thinner
covering density than that 
of the $A_d^*$ lattice in the large-$d$ limit [cf. (\ref{A_n^*})],
but it is not known whether this bound becomes sharp in the large-$d$
limit.

The best lower bound on the covering density
is given by
\begin{equation}
\theta \ge \tau_d.
\label{cover-low}
\end{equation}
To define $\tau_d$, let $S$ be a regular simplex with edge length
equal to two. Spheres of radius $\sqrt{2d/(d+1)}$ centered
at the vertices of $S$ just cover $S$. The quantity
$\tau_d$ is the ratio of the sums of the intersections
of these spheres with $S$ to the volume of $S$.
Thus, in the large-$d$ limit, $\tau \rightarrow d/e^{3/2}$.

\subsection{Quantizer Problem}
\label{quantizer}

\begin{figure}
 \centerline{ \includegraphics[width=2.2in,keepaspectratio]{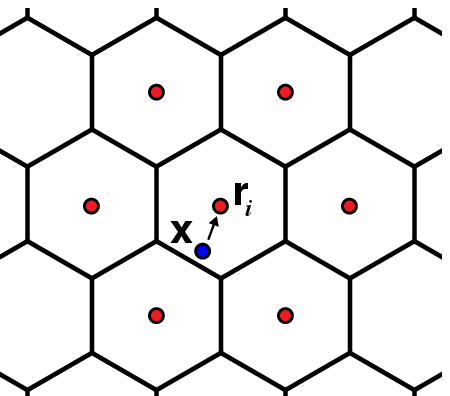}
\hspace{0.55in}\includegraphics[width=2.in,keepaspectratio,clip=]{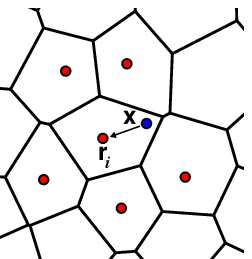}}
\caption{(Color online) Examples of two-dimensional quantizers. Any point $\bf x$
is quantized (``rounded-off") to the nearest point ${\bf r}_i$. Left panel:
Triangular lattice (best quantizer in $\mathbb{R}^2$ \cite{Co93}). Right panel: Irregular point process.}
\label{tri-RSA}
\end{figure}

Consider a point process ${\cal P}$  in $\mathbb{R}^d$ with configuration ${\bf r}_1,{\bf r}_2,\ldots, {\bf r}_N$.
A $d$-dimensional {\it quantizer} is device that takes as an input  a point 
at position ${\bf x}$ in $\mathbb{R}^d$ generated
from some probability density function $p({\bf x})$ and outputs  the nearest point
${\bf r}_i$ of the point process to $\bf x$. Equivalently, if the input $\bf x$ belongs
to the Voronoi cell ${\cal V}({\bf r}_i)$, the output is ${\bf r}_i$
(see Fig.~\ref{tri-RSA})
For simplicity, we assume that $\bf x$ is uniformly distributed
over a large ball in $\mathbb{R}^d$ containing the $N$ points of
the point process. One attempts to choose the configuration 
${\bf r}_1,{\bf r}_2,\ldots, {\bf r}_N$ of the point process 
to minimize the mean squared error, i.e., the expected
value of $|{\bf x}- {\bf r}_i|^2$. Specifically, the quantizer
problem is to choose the $N$-point configuration
so as to minimize the {\it scaled dimensionless error} 
(sometimes called the {\it distortion}) \cite{Co93}
\begin{equation}
{\cal G}= \frac{1}{d} \langle R^2 \rangle,
\label{gamma1}
\end{equation}
where 
\begin{equation}
\langle R^2 \rangle= \frac{\displaystyle \lim_{N \rightarrow \infty} \frac{1}{N} \sum_{i=1}^N 
 \int_{{\cal V}({\bf r}_i)} |{\bf x}-{\bf r}_i|^2 d{\bf x}}{\displaystyle \langle V({\cal V}) \rangle ^{1+\frac{2}{d}}}.
\end{equation}
is a dimensionless error,
\begin{equation}
\langle V({\cal V}) \rangle =\displaystyle \left[\lim_{N \rightarrow \infty}\frac{1}{N} \sum_{i=1}^N V({\cal V}
({\bf r}_i))\right]
\end{equation}
is the expected volume of a Voronoi cell, and $V({\cal V}({\bf r}_i))$
is the volume of the $i$th Voronoi cell.
The scaling factor $1/d$ is included to compare the second moments appropriately across dimensions. We will denote the minimal scaled dimensionless error by ${\cal G}_{min}$.
Note that since there is one point per Voronoi cell, the number density
$\rho$ is set equal to unity.

If all of the Voronoi cells are congruent (as they are in the case
of all lattices and some periodic point processes), we have the simpler expression
\begin{equation}
{\cal G} = \frac{\frac{\displaystyle 1}{\displaystyle d}
 \int_{{\cal V}} |{\bf x}|^2 d{\bf x}}{V({\cal V})^{1+\frac{2}{d}}}, 
\label{gamma2}
\end{equation}
where the centroid of the Voronoi cell $\cal V$ is the origin of the coordinate system.
The {\it lattice quantizer problem} is to find the lattice for which $\cal G$, given
by (\ref{gamma2}), is minimum. Thus, $\cal G$ can be interpreted as the
scaled, dimensionless {\it second moment of inertia} of the Voronoi cell.


\begin{table}[bthp]
\caption{Best known solutions to the 
quantizer problem in selected dimensions. Values
reported are taken from Conway and Sloane \cite{Co93}, except
for $d=9$ and 10, which were obtained from  Agrell and Eriksson \cite{Ag98}. It is only in one and two dimensions that these
solutions have been proved to be globally optimal \cite{Co93}.}
\label{quant}
\renewcommand{\baselinestretch}{1.2} \small \normalsize
\centering
\begin{tabular} {|c|c|c|}
\multicolumn{3}{c}{~} \\\hline
 {Dimension, $d$} & Quantizer & Scaled Error, $\cal G$ \\ \hline
1 &  $A_1^*= \mathbb{Z}$ & 0.083333 \\

2 & $A_2^*\equiv A_2$  & 0.080188 \\

3 &  $A_3^*\equiv D_3^*$  & 0.078543 \\

4  & $D_4^*\equiv D_4$  & 0.076603 \\
5  & $D_5*$ & 0.075625 \\
6  & $E_6^*$  & 0.074244  \\
7  & $E_7^{*}$ & 0.073116 \\
8  & $E_8^*= E_8$  & 0.071682 \\
9 &  $L^{AE}_9$ & 0.071626 \\
10&  $D^+_{10}$ & 0.070814  \\
12  & $K_{12}^*\equiv K_{12}$ & 0.070100 \\
16 & $\Lambda_{16}^*\equiv \Lambda_{16}$  & 0.068299 \\
24 & $\Lambda_{24}^*= \Lambda_{24}$  & 0.065771 \\ \hline
\end{tabular}
\end{table}

The best known quantizers in any dimension $d$ are
usually lattices that are the duals of the densest known packings  (see
the discussion in Sec. \ref{sphere} and Ref.~\cite{Co93}), except in dimensions
9 and 10, where the best solutions are still lattices,
but are not duals of the densest lattice packings in those
dimensions. Although the best known solutions of the 
quantizer and covering problems are the same in the first three space
dimensions, they are generally different for $d \ge 4$ \cite{Co93,Sc06}.
Zador \cite{Za82} has derived upper and lower bounds on ${\cal G}_{min}$.
Conway and Sloane \cite{Co93} have obtained conjectural lower bounds
on ${\cal G}_{min}$. We defer the discussion of these bounds to Sec.~\ref{results-quant},
where we derive sharper upper bounds on ${\cal G}_{min}$, among other results.
Table~\ref{quant} provides the best known solutions to
the quantizer problem in selected dimensions.

\subsection{Comparison of the Four Problems}
\label{four}

Table \ref{comparison} lists the best known solutions of the quantizer, covering and number-variance
problems and sphere-packing problems  in  $\mathbb{R}^d$  for selected $d$.
It is seen that in the first two space dimensions, the best known solutions for
each of these four problems are identical to one another.
For $d=3$, the densest sphere packing is the $D_3$ or, equivalently, $A_3$ lattice,
which is the dual lattice associated with the best known solutions to the quantizer, covering and number-variance
problems, which is the $A_3^*$ lattice. Thus, for the first three space dimensions, 
the best known solutions for each of the four problems are lattices, and either they
are identical to one another or are duals of one another. However, such  relationships 
may or may not  exist for $d \ge 4$, depending on the peculiarities
of the dimensions involved.

\begin{table}[bthp]
\caption{Comparison of the best known solutions to the quantizer,
covering, number variance and sphere packing problems, as obtained from the previous four tables. Recall that the $E_8$ and $\Lambda_{24}$ lattices
are self-dual lattices.}
\label{comparison}
\renewcommand{\baselinestretch}{1.2} \small \normalsize
\centering
\begin{tabular} {|c|c|c|c|c|}
\multicolumn{5}{c}{~} \\\hline
 {Dimension, $d$} & {Quantizer}& Covering& Variance & Packing\\ \hline
1 &  $A_1^*=\mathbb{Z}$ & $A_1^*=\mathbb{Z}$  & $A_1^*= \mathbb{Z}$ & $A_1^*= \mathbb{Z}$ \\

2 & $A_2^*\equiv A_2$  & $A_2^*\equiv A_2$  & $A_2^*\equiv A_2$ & $A_2^*\equiv A_2$ \\

3 &  $A_3^*\equiv D_3^*$  & $A_3^*\equiv D_3^*$  & $A_3^*\equiv D_3^*$ & $A_3\equiv D_3$ \\

4 &  $D_4^*\equiv D_4$ & $A_4^*$  & $D_4^*\equiv D_4$ & $D_4^*\equiv D_4$ \\

5 &  $D_5^{*}$  & $A_5^*$  & $\Lambda_5^{2*}$ & $D_5$ \\

6 &  $E_6^*$ & $L_6^{c1}$  & $E_6^*$ & $E_6$ \\ 
7 &  $E_7^*$ & $L_7^c$  & $\Lambda_7^{3*}$ & $E_7$ \\ 
8 &  $E_8$ & $L_8^c$  & $E_8$ & $E_8$ \\ 
9 &  $L^{AE}_9$ & $A^5_9$  & $\Lambda^*_9$  & $\Lambda_9$ \\ 
10&  $D^+_{10}$ & $A^*_{10}$   & $\Lambda^*_{10}$  & $P_{10c}$ \\ 
12&  $K_{12}$ &  $A^*_{12}$  & $\Lambda^{max*}_{12}$ & $\Lambda^{max}_{12}$ \\ 
16&  $\Lambda^*_{16}$ & $A^*_{16}$   & $\Lambda^*_{16}$& $\Lambda_{16}$ \\ 
24&  $\Lambda_{24}$ & $\Lambda_{24}$  & $\Lambda_{24}$ & $\Lambda_{24}$ 
\\ \hline
\end{tabular}
\end{table}

There is a fundamental difference between the nature of the interactions
for the sphere-packing problem
and those for the other three problems.
Any sphere packing (optimal or not) consisting of
nonoverlapping spheres of diameter $D$ is described by
a short-ranged pair potential that is zero whenever the spheres
do not overlap (when the pair separation distance is greater than $D$)
and is infinite whenever the pair separation distance
is less than $D$. By contrast, the
other three problems are described by ``soft" bounded interactions.
In particular, we have seen that the number variance is specified by a bounded
repulsive pair potential with compact support [cf. (\ref{volavginterp})]. We will see in the
subsequent sections that the covering and quantizer problems
are described by many-particle bounded interactions 
but of the more general form (\ref{full}), which 
involves single-body, two-body,
three-body, and  higher-body interactions.


One simple reason why the optimal solutions of the sphere-packing and number-variance
problems are related to one another either directly or via their dual solutions
in the first three space dimensions is that they both involve short-ranged 
repulsive pair interactions only. The reader is referred to Refs. \cite{To03a} 
and \cite{To08a} for a comprehensive explanation. The reasons
why the optimal solutions to these two problems are sometimes
the optimal solutions for the covering and quantizer problems
will become apparent in the subsequent sections.
The explanation for why the optimal covering and quantizer solutions
are generally different for $d \ge 4$ is discussed in Sec.~\ref{reform}.
We note that the Leech lattice $\Lambda_{24}$
for $d=24$ is an exceptional case in that it provides the optimal
solution to all four different problems. The remarkably
high degree of symmetry possessed by this self-dual lattice \cite{Co09}
accounts for this unique property. The only other dimensions
where all four optimal solutions are the same  are $d=1$ and $d=2$.

\section{Nearest-Neighbor Functions}
\label{near}

\subsection{Definitions}

We recall the definition of the ``void" nearest-neighbor probability
density function $H_V(R)$ \cite{To02a}: 
\begin{eqnarray}
\begin{array}{ccp{3.7in}}
H_V(R)\, d R & = & Probability that a  point of the point process
lies at a distance between $R$ and $R + d R$ from a randomly
chosen point in $\mathbb{R}^d$. 
\end{array} 
\label{def-Hv}
\end{eqnarray}
The ``void" exclusion probability $E_V(R)$ is the {\it complementary cumulative distribution
function} associated with $H_V(R)$:
\begin{eqnarray}
E_V(R) = \int_R^\infty H_V(x) dx,
\label{Ev-cum}
\end{eqnarray}
and hence is a monotonically decreasing function of $R$ \cite{To02a}.
Thus, $E_V(R)$ has the following probabilistic interpretation:
\begin{eqnarray}
\begin{array}{ccp{3.7in}}
E_V(R) & = & Probability of finding a randomly placed spherical cavity
of radius $R$ empty of any points. 
\end{array}
\label{def-Ev}
\end{eqnarray}
There is another interpretation of $E_V$ that involves circumscribing spheres of
radius $R$ around each point in a realization of  the point process. It immediately
follows that $E_V(R)$ is the {\it expected} fraction of space not covered by these circumscribing spheres. Differentiating (\ref{Ev-cum}) with
 respect to $R$ gives
\begin{eqnarray}
H_V(R) = -\frac{\partial E_V}{\partial R}.
\label{derv-Ev}
\end{eqnarray}
Note that these void quantities are different from the ``particle" nearest-neighbor
functions \cite{To90c,To95a,To08c} in which  the sphere of radius $R$ is centered at an actual point of the
point process (as opposed to an arbitrary point in the space).

It is useful to introduce the ``conditional" nearest-neighbor
function $G_V(r)$ \cite{Re59,To90c}, which is defined in terms of  $H_V(r)$ and $E_V(r)$ as
follows:
\begin{eqnarray}
H_{V}(R) = \rho s_{1}(R) G_{V}(R) E_{V}(R),
\label{def-Hv2}
\end{eqnarray}
where $s_1(R)$ is the surface area of a 
$d$-dimensional sphere of radius $R$ [cf. (\ref{area-sph})].
Thus, we have the following interpretation of the conditional function:
\begin{eqnarray}
\begin{array}{ccp{3.1in}}
\rho s_{1}(R) G_{V}(R) \;d R & = &
Given that a spherical cavity of radius $R$ centered
at an arbitrary point in the space is
empty of any points of the point process, the probability of finding
a point in the spherical shell of volume
$s_{1}(R)\,d R$ surrounding the arbitrary point.
\end{array}
\label{def-Gv}
\end{eqnarray}
Therefore, it follows from (\ref{derv-Ev}) and (\ref{def-Hv2})
that the exclusion probability can be expressed in terms of $G_V$ via
the relation
\begin{eqnarray}
E_V(R)=\exp\left[-\rho s_1(1) \int_0^R x^{d-1} G_V(x) \;d x\right].
\label{EV-GV}
\end{eqnarray}
It is clear that the void functions have the following behaviors at the origin for $d\ge 2$ \cite{hv}:
\begin{eqnarray}
E_V(0)=1, \qquad H_V(0)=0, \qquad G_V(0)=1.
\label{origin}
\end{eqnarray}

Moments of the nearest-neighbor function $H_V(R)$ arise in rigorous bounds for transport properties
of random media \cite{To02a}. The $n$th moment of $H_V(R)$ is defined as
\begin{eqnarray}
\langle R^n \rangle = \int_{0}^{\infty} R^n H_V(R) \, dR =n \int_0^\infty R^{n-1} E_V(R) \,dR\label{nth-mom}.
\end{eqnarray}

\subsection{Series Representations}

The void functions can be expressed as infinite series
whose terms are integrals over the $n$-particle density functions \cite{To02a,To90c}.
For example, the void exclusion probability functions for a translationally
invariant point process are respectively given by
\begin{eqnarray}
E_V(R) = 1+ \sum^{\infty}_{k=1} (-1)^{k} \frac{\rho^k}{k!}
 \int_{{\mathbb R}^d} g_{k}({\bf r}_1,\ldots,{\bf r}_k) \prod_{j=1}^k \Theta(R-|{\bf x} - {\bf r}_{j}|) d{\bf r}_j,
\label{Ev-series}
\end{eqnarray}
where $g_n$ is the $n$-particle correlation function and 
$\Theta(x)$ is the Heaviside step function defined by (\ref{heaviside}).  The corresponding series for $H_V(R)$  is obtained from the series above using (\ref{derv-Ev}). 

Note that the series (\ref{Ev-series}) can be rewritten in terms
of intersection volumes of spheres:
\begin{eqnarray}
E_V(R) = 1+ \sum^{\infty}_{k=1} (-1)^{k} \frac{\rho^k}{k!}
 \int_{{\mathbb R}^d} g_{k}({\bf r}_1,\ldots,{\bf r}_k)\, 
v_k^{\mbox{\scriptsize int}}({\bf r}_1,\ldots,{\bf r}_k;R)\,  d{\bf r}_1\cdots d{\bf r}_k,
\label{F-series}
\end{eqnarray}
where
\begin{equation}
v_n^{\mbox{\scriptsize int}} ({\bf r}_{1}, \ldots, {\bf r}_{n}; R) =
\int \; d{\bf x} \; \prod_{j = 1}^{n} \; \Theta(R - |{\bf x}-{\bf r}_j|)
\label{inter-vol}
\end{equation}
is the intersection volume  of $n$ equal spheres
of radius $R$ centered at positions ${\bf r}_1, \ldots, {\bf r}_n$.
Observe that $v_2^{\mbox{\scriptsize int}}(r;R))=v_1(R)\alpha(r;R)$, where
$v_1(R)$ is the volume of a sphere of radius $R$ [cf. (\ref{v1})]
and $\alpha(r;R)$ is the scaled intersection volume [cf. (\ref{alpha})] and (\ref{series})].

In the special case of a Poisson point distribution, $g_n=1$ for all $n$, and hence
(\ref{F-series}) immediately yields the well-known exact result
for such a spatially uncorrelated point process
\begin{equation}
E_V(R)=\exp(-\rho v_1(R)).
\label{Poisson}
\end{equation}
The use of this relation with definition (\ref{derv-Ev}) gives
\begin{equation}
H_V(R)=\rho s_1(R) \exp(-\rho v_1(R)).
\end{equation}

For a single realization of $N$ points within a large volume $V$ in $\mathbb{R}^d$, we have
\begin{equation}
E_V(R) = 1 - \rho v_1(R)+\frac{1}{V}\sum_{i <j} v_2^{\mbox{\scriptsize int}}(r_{ij};R) -\frac{1}{V}\sum_{i<j < k} 
v_3^{\mbox{\scriptsize int}}(r_{ij},r_{ik},r_{jk};R) - \cdots
\label{EV}
\end{equation}
This formula assumes that $N$ is sufficiently large so that boundary effects can be neglected.
The second term in (\ref{EV}) $\rho v_1(R)= \sum_{i=1}^N v_1(R)/V$ can be interpreted
as a sum over one-body terms, which is independent of the point configuration. Clearly,
the ($n+1$)th term in (\ref{EV}) can be interpreted as a sum over intrinsic $n$-body 
interactions,namely, $v_n^{\mbox{\scriptsize int}}$. Thus, except for the trivial constant of unity (the first term),
$E_V(R)$ can be regarded to be a  many-body
potential of the general form (\ref{full}), which heretofore was not observed. 

\subsection{Rigorous Bounds on the Nearest-Neighbor Functions}
\label{bounds-E}

Upper and lower bounds on
 the so-called  {\it canonical $n$-point correlation function} 
$H_{n} ({\bf x}^{m}; {\bf x}^{p-m}; {\bf r}^{q})$ (with $n=p+q$ and $m \le p$) for 
point processes in $\mathbb{R}^d$ have been found \cite{To02a,To86i}.  Since the void 
exclusion probability and nearest-neighbor probability density function are just special cases of $H_{n}$,
 then we also have strict bounds on them for such models.
 Let $X$ represent either $E_V$ or $H_V$
 and $X^{(k)}$ represent the $k$th term of the series
for these functions. Furthermore, let
\begin{eqnarray}
W^{\ell}  =  \sum_{k=0}^{\ell} (-1)^{k} X^{(k)} 
\end{eqnarray}
be the partial sum.  Then it follows   that
 for any of the exclusion probabilities or nearest-neighbor
 probability density functions, we have the bounds
\begin{eqnarray}
X & \leq & W^{\ell}, \qquad \mbox{for $\ell$ even}
\mbox{} \nonumber \\
X & \geq & W^{\ell}, \qquad \mbox{for $\ell$ odd}.
\end{eqnarray}

Application of the aforementioned inequalities yield the 
first three successive bounds on the nonnegative exclusion probability:
\begin{eqnarray}
E_V(R) &\le& 1 \\
E_V(R) &\ge& 1 -\rho v_1(R) \label{EV-bound1}\\
E_V(R)  &\le&  1 -\rho v_1(R) +\frac{\rho^2}{2} s_1(1)\int_0^{2R} x^{d-1} v_2^{int}(x;R) g_2(x) d x
\label{EV-bound2},
\end{eqnarray}
where $v_2^{int}(x; R) = v_1(R)\alpha(x;R)$ is the intersection volume of two $d$-dimensional spheres 
of radius $R$ whose centers are separated by the distance $x$ and $\alpha(x;R)$
is the scaled intersection volume given by (\ref{alpha}).
The corresponding first two nontrivial bounds on the 
nonnegative pore-size density function $H_V(R)$
are as follows:
\begin{eqnarray}
H_V(R) &\le& \rho s_1(R) \label{HV-bound1}\\
H_V(R)  &\ge&  \rho s_1(R) -
\frac{\rho^2}{2} s_1(1)\int_0^{2R} x^{d-1} s_2^{int}(x;R) g_2(x) \;d x,
\label{HV-bound2}
\end{eqnarray}
where $s_2^{int}(x;R)\equiv \partial v_2^{int}(x;R)/\partial R$
is the surface area of the intersection volume $v_2^{int}(x;R)$.
Bounds on the conditional
function $G_V(r)$ follow by combining the bounds above on $E_V(r)$ and $H_V(r)$ and definition
(\ref{def-Hv2}). For example, the following bounds have been found \cite{To08c}:
\begin{eqnarray}
G_V(R) \le \frac{1}{1-\rho v_1(R)}
\label{GV-bound1}
\end{eqnarray}
and
\begin{eqnarray}
G_V(R) \ge \frac{1 - \frac{\rho}{s_1(R)} s_1(1)\int_0^{2R} x^{d-1} s_2^{int}(x;R) g_2(x) \;d x}
{1 -\rho v_1(R) +\frac{\rho^2}{2} s_1(1)\int_0^{2R} x^{d-1} v_2^{int}(x;R) g_2(x) \;d x},
\label{GV-bound2}
\end{eqnarray}
which should only be applied for $R$ such that $G_V(R)$ remains positive.

\subsection{Truncation of the Series Expansions for Nearest-Neighbor Functions for Packings}

For congruent sphere packings of diameter $D$ 
at packing density $\phi$, the infinite series expansion
for $E_V(R)$ [cf. (\ref{Ev-series})] will truncate after a finite number of terms
for a bounded value of the radius $R$. A spherical region
of radius $R$ centered at an arbitrary point in the space
exterior to the spheres can contain at most $n_{max}$ sphere centers.
Therefore, series truncates after $n_{max}+1$ terms, i.e.,
\begin{eqnarray}
E_V(R) = 1+ \sum^{n_{max}}_{k=1} (-1)^{k} \frac{\rho^k}{k!}
 \int_{{\mathbb R}^d} g_{k}({\bf r}_1,\ldots,{\bf r}_k) \prod_{j=1}^k \Theta(R-|{\bf x} - {\bf r}_{j}|) d{\bf r}_j,
\label{Ev-series2}
\end{eqnarray}

For a spherical region of radius $R=D/\sqrt{3}$, $n_{max}=2$ and hence we have the exact expression that applies for $0 \le R \le D/\sqrt{3}$
\begin{eqnarray}
E_V(R)  &=&  1 -\rho v_1(R) +\frac{\rho^2}{2} s_1(1)\int_D^{2R} x^{d-1} v_2^{int}(x;R) g_2(x) d x\nonumber \\
&=& 1- 2^d \phi \left(\frac{R}{D}\right)^d + \frac{2^{d-1} d \phi^2}{D^d} \left(\frac{R}{D}\right)^d 
\int_D^{2R} x^{d-1} \alpha(x;R) g_2(x) d x.
\label{EV-packing}
\end{eqnarray}
For any sphere packing for which (\ref{period}) applies
such that $r_2/r_1 > 2/\sqrt{3}$, we have upon use of (\ref{EV-packing}) 
the exact result
\begin{equation}
E_V(R) = \left\{\begin{array}{ll}
1- 2^d \phi \left(\frac{R}{D}\right)^d,   & 0 \le R \le D/2\\
1- 2^d \phi \left[ 1 -\frac{Z}{2}  \alpha(D;R)\right] 
\left(\frac{R}{D}\right)^d, & D/2 \le R \ge D/\sqrt{3},
\end{array}
\right.
\label{EV-packing2}
\end{equation}
where we have used the fact that $r_1=D$. Importantly, this relation applies
to the densest known lattice packings of spheres, at least for
dimensions in the range $1 \le d \le 24$.

\section{Reformulation of the Covering and Quantizer Problems}
\label{reform}

\subsection{Reformulations}

Now we can reformulate the covering and quantizers problems in terms
of the void exclusion probability. In particular, the covering
problem asks for the point process in $\mathbb{R}^d$ at unit density
($\rho=1$) that minimizes the support of the radial function $E_V(R)$. We define  ${\cal R}_c^{min}$ the 
smallest possible value of the covering radius ${\cal R}_c$ among all point processes for which $E_V(R)=0$,
which we call  the {\it minimal covering radius}.
This is indeed a special ground state in which the ``energy"  is
identically zero (i.e., $E_V({\cal R}_c^{min})=0$) \cite{ground}.
Depending on the space dimension $d$,
this special ground state  will involve one-body interactions, the first two terms
of (\ref{EV}), one- and two-body interactions, the first three terms
of (\ref{EV}), one-, two- and three-body interactions, the first four terms
of (\ref{EV}), etc., and will truncate at some particular level, provided
that $E_V(R)$ for the point process has compact support. The minimal
covering radius ${\cal R}_c^{min}$ increases with the space dimension $d$
and, generally speaking, the highest-order  $n$-body interaction
required to fully characterize the associated $E_V(R)$ increases
with $d$. Note that for a particular point process, twice the covering radius 
$2{\cal R}_c$ can be viewed as the ``effective interaction range" between any pair of points,
since the intersection volume $v_2^{\mbox{\scriptsize int}}(r_{ij};R)$, which appears in expression (\ref{EV}) for $E_V(R)$, is exactly zero for any pair separation
$r_{ij} > 2 {\cal R}_c$; moreover, for such pair separations
$ v_n^{\mbox{\scriptsize int}}$ can be written purely in terms
of the lower-order intersection volume $ v_{n-1}^{\mbox{\scriptsize int}}$.
Therefore, because $v_2^{\mbox{\scriptsize int}} \ge  v_n^{\mbox{\scriptsize int}}$ for $n\ge 3$, the
effective interaction range between any $n$ points for $n\ge 3$ is still
given by $2{\cal R}_c$.

The quantizer problem asks for the point process in $\mathbb{R}^d$ at unit density
that minimizes the scaled average squared error ${\cal G}$ defined as
\begin{equation}
{\cal G} = \frac{1}{d}\langle R^2 \rangle=\frac{1}{d} \int_0^\infty 
R^2 H_V(R) dR=\frac{2}{d} \int_0^\infty R E_V(R) dR.
\label{G}
\end{equation}
We will call the minimal error ${\cal G}_{min}$. Thus, we seek the ground
state of the many-body interactions that are involved upon substitution
of (\ref{EV}) into (\ref{G}). Again, depending on the dimension, 
this many-body energy  will truncate at some particular level, provided
that $E_V(R)$ has compact support. Again, as $d$ increases, successively higher-order
interactions in the expression (\ref{EV}) must be incorporated
to completely characterize $E_V(R)$.

\subsection{Explicit Calculations for Some Common Lattices Using These Reformulations}

It is instructive to express explicitly the void exclusion probabilities
for some common lattices and use these functions to evaluate
explicitly their corresponding covering densities and scaled average squared errors
in the first three space dimensions.

In the simplest case of one dimension, the series expansion
for $E_V(R)$ [cf. (\ref{EV})] for the integer lattice
$\mathbb{Z}$ truncates after only one-body terms. At unit number
density ($\rho=1$), it is trivial to show that
\begin{equation}
E_V(R) = \left\{\begin{array}{ll}
1- v_1(R),  & 0 \le R \le {\cal R}_c\\
0, & R \ge {\cal R}_c,
\end{array}
\right.
\label{integer}
\end{equation}
where the covering radius ${\cal R}_c=1/2$, $v_1(R)=2R$, 
and the nearest-neighbor distance from a lattice point
is unity. Using the definitions   (\ref{covering}) and (\ref{G})
in combination with (\ref{integer}) yield
the covering density and scaled average squared error
, respectively, for the optimal integer lattice:
\begin{equation}
\theta=1,
\end{equation}
\begin{equation}
{\cal G} =\frac{1}{12}=0.083333\ldots.
\end{equation}

\begin{figure}[bthp]
\centerline{ \includegraphics[width=3.5in]{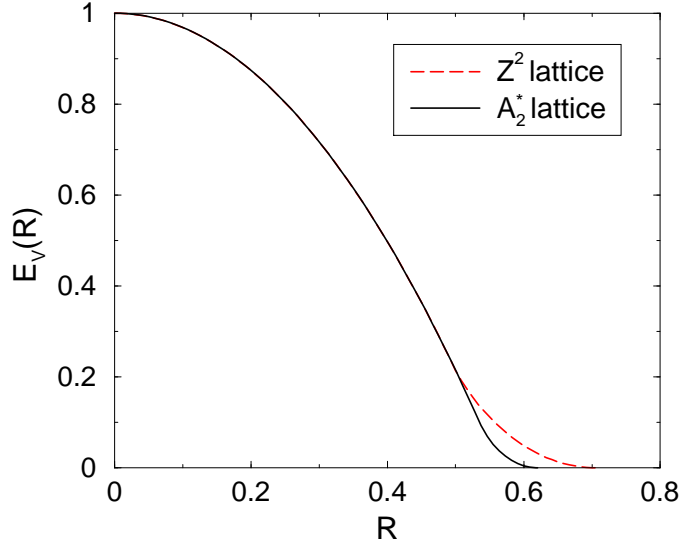}}
\caption{ (Color online) The void exclusion probability $E_V(R)$ for the
$\mathbb{Z}^2$ (square) and $A_2\equiv A_2^*$ (triangular)
lattice have support up to the covering radii ${\cal R}_c=\sqrt{2}/2=0.7071\ldots$ and ${\cal R}_c=\sqrt{2}/3^{3/4}=0.6204\ldots$, respectively, at unit number density ($\rho=1$). }
\label{2D-lattices}
\end{figure}

Let us now determine the void exclusion probabilities for 
the $\mathbb{Z}^2$ (square) and $A_2\equiv A_2^*$ (triangular) and lattices
for $R$ up to their respective covering radii for which $E_V(R)=0$.
For these lattices, the series expression (\ref{EV}) for $E_V(r)$ 
truncates after two-body terms. For the square and triangular lattices at $\rho=1$,
\begin{equation}
E_V(R) = \left\{\begin{array}{ll}
1- v_1(R),   & 0 \le R \le r_1/2,\\
1- v_1(R) + 2v_2^{\mbox{\scriptsize int}}(r_1;R),   &  r_1/2 \le R \le {\cal R}_c, \\
0, & R \ge {\cal R}_c
\end{array}
\right.
\label{square}
\end{equation} 
where $r_1$ is the nearest-neighbor distance from a lattice point
and the covering radius  ${\cal R}_c$, equal to one half of the 
next-nearest-neighbor distance, which we will denote by $r_2$. 
For the square and triangular lattices at $\rho=1$, 
$r_1=1$ and ${\cal R}_c=\sqrt{2}/2=0.7071\ldots$, and $r_1=\sqrt{2}/3^{1/4}=1.0745\ldots$ and 
${\cal R}_c=\sqrt{2}/3^{3/4}=0.6204\ldots$
respectively. Figure \ref{2D-lattices} provides plots of $E_V(R)$ for
these two $d=2$ lattices.
Employing the definitions   (\ref{covering}) and (\ref{G})
in combination with (\ref{square}) provide
the covering density and scaled average squared error, respectively, for the $\mathbb{Z}^2$ lattice:
\begin{equation}
\theta=\frac{\pi}{2}=1.57079\ldots,
\end{equation}
\begin{equation}
{\cal G} =\frac{1}{12}=0.083333\ldots.
\end{equation}
Similarly, the corresponding equations  for the $A_2$
lattice yields the covering density and scaled average squared error
, respectively, for this optimal structure:
\begin{equation}
\theta=\frac{2 \pi}{3^{1/3}}=1.20919\ldots,
\end{equation}
\begin{equation}
{\cal G} =\frac{5}{36\sqrt{3}}=0.08018\ldots.
\end{equation}

It is also useful to express explicitly the void exclusion probabilities
for the $\mathbb{Z}^3$ (simple cubic) and $A_3^*$ (bcc) lattices
for $R$ up to their respective covering radii for which $E_V(R)=0$.
It turns out that calculating $E_V(R)$ in the case
of the simple-cubic lattice is more complicated than
that for the bcc lattice because the former involves up
through four-body terms, i.e., $v_4^{\mbox{\scriptsize int}}$.
However, the symmetry of the geometry for $\mathbb{Z}^3$ enables one to 
express $v_4^{\mbox{\scriptsize int}}$ purely in terms
of  $v_2^{\mbox{\scriptsize int}}$ and $v_3^{\mbox{\scriptsize int}}$.
The calculation of $E_V(R)$ does not involve four-body
terms.  For the simple cubic lattice at $\rho=1$,
\begin{equation}
 E_V(R) = \left\{\begin{array}{ll}
1- v_1(R),   & 0 \le R \le r_1/2,\\
1- v_1(R) + 3v_2^{\mbox{\scriptsize int}}(r_1;R),   &  r_1/2 \le R \le r_2/2,\\
1- v_1(R) + 3v_2^{\mbox{\scriptsize int}}(r_1;R)  +
3v_2^{\mbox{\scriptsize int}}(r_2;R) -
6v_3^{\mbox{\scriptsize int}}(r_1,r_1,r_2;R), &  r_2/2 \le R \le {\cal R}_c,\\
0, & R \ge {\cal R}_c,
\end{array}
\right.
\label{sc}
\end{equation}
where $r_1=1$, $r_2=\sqrt{2}=1.4142\ldots$, ${\cal R}_c=\sqrt{3}/2=0.8660\ldots$ and 
$v_3^{\mbox{\scriptsize int}}(r,s,t;R)$ is explicitly
given by (\ref{v3int}) in Appendix \ref{v3-int} for triangles of side lengths $r$, $s$ and $t$.
Here we have used the fact that for $r_2/2 \le R \le {\cal R}_c$,
$v_4^{\mbox{\scriptsize int}}(r_1,r_1,r_2,r_2)= 
6v_3^{\mbox{\scriptsize int}}(r_1,r_1,r_2)- 3 v_2^{\mbox{\scriptsize int}}(r_2)$.
Using the definitions   (\ref{covering}) and (\ref{G})
in combination with (\ref{sc}) yield
the covering density and scaled average squared error, respectively, for the  $\mathbb{Z}^3$ lattice:
\begin{equation}
\theta=\frac{\pi\sqrt{3}}{2}=2.72069\ldots,
\end{equation}
\begin{equation}
{\cal G} =\frac{1}{12}=0.083333\ldots.
\end{equation}

\begin{figure}[bthp]
\centerline{ \includegraphics[width=3.5in]{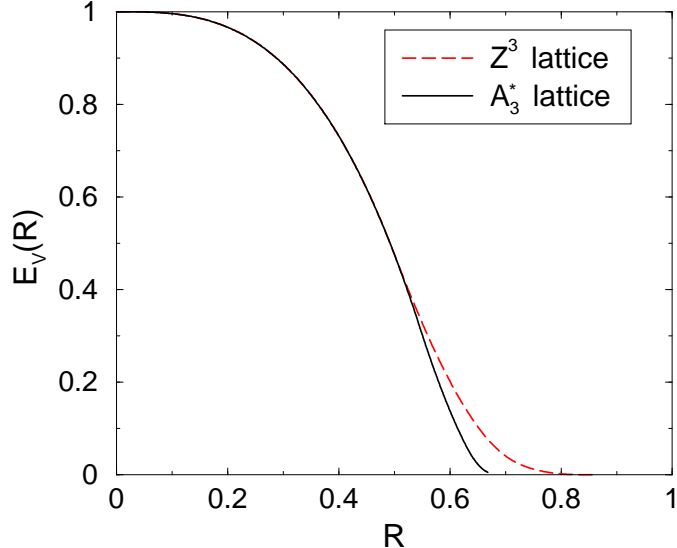}}
\caption{ (Color online) The void exclusion probability $E_V(R)$ for the $\mathbb{Z}^3$ (simple cubic) lattice and $A_3^*$ (bcc) lattice have 
support up to the covering radii  ${\cal R}_c=\sqrt{3}/2=0.8660\ldots$ and ${\cal R}_c=\sqrt{5}/2^{5/3}=0.7043\ldots$, respectively, at unit number density ($\rho=1$). }
\label{3D-lattices}
\end{figure}

For the bcc lattice at $\rho=1$,
\begin{equation}
 E_V(R) = \left\{\begin{array}{ll}
1- v_1(R),   & 0 \le R \le r_1/2,\\
1- v_1(R) + 4v_2^{\mbox{\scriptsize int}}(r_1;R),   &  r_1/2 \le R \le r_2/2,\\
1- v_1(R) + 4v_2^{\mbox{\scriptsize int}}(r_1;R)  +3v_2^{\mbox{\scriptsize int}}(r_2;R),
  &  r_2/2 \le R \le R_T,\\
1- v_1(R) + 4v_2^{\mbox{\scriptsize int}}(r_1;R)  +3v_2^{\mbox{\scriptsize int}}(r_2;R) -12v_3^{\mbox{\scriptsize int}}(r_1,r_1,r_2;R), &  R_T \le R \le {\cal R}_c,\\
0, & R \ge {\cal R}_c,
\end{array}
\right.
\label{bcc}
\end{equation}
where $r_1=\sqrt{3}/4^{1/3}=1.0911\ldots$, $r_2=4^{1/6}=1.2599\ldots$, 
${\cal R}_c=\sqrt{5}/2^{5/3}=0.7043\ldots$,
and $R_T=3/2^{13/6}=0.6681\ldots$ is the circumradius of a triangle
of side lengths $r_1$, $r_1$ and $r_2$, the general
expression of which is given by (\ref{circum}) in Appendix \ref{v3-int}.
Employing the definitions   (\ref{covering}) and (\ref{G})
in combination with (\ref{bcc}) provide
the covering density and scaled mean squared error, respectively, for the $\mathbb{A}^*_3$ lattice:
\begin{equation}
\theta=\frac{\pi \cdot 5^{3/2}}{24}=1.46350\ldots,
\end{equation}
\begin{equation}
{\cal G} =\frac{19}{192 \cdot 2^{1/3}}=0.078543\ldots.
\end{equation}
Figure \ref{3D-lattices} provides plots of  $E_V(R)$ for
the simple cubic and bcc three-dimensional lattices.

\subsection{Remarks About Higher Dimensions}

We see that the best known solutions to the covering and quantizer
problems are identical for the first three space dimensions.
However, there is no reason to expect that the optimal solutions
for these two problems to be the same in higher dimensions, except
for $d=24$ for reasons mentioned in Sec.~\ref{four}.
Although both problems involve ground states associated with the ``many-body
interaction" function $E_V(R)$,
such that it possesses compact support for finite $d$, the precise
shape of the function $E_V(R)$ for a particular point configuration 
is crucial in determining its first moment
or quantizer error. By contrast, the best covering 
seeks to find the point configuration that minimizes the support
of $E_V(R)$ without regard to its shape.

\section{Results for the Covering Problem}
\label{results-cover}

Saturated sphere packings in $\mathbb{R}^d$ should provide
relatively thin coverings. Surrounding every sphere of diameter $D$ in any saturated packing
of congruent spheres in $\mathbb{R}^d$ at packing density $\phi_s$ by spheres of radius $D$ provides
a covering of $\mathbb{R}^d$, and thus the associated covering density $\theta_s$
is given by
\begin{equation}
\theta_s= \rho _s v_1(D)= 2^d \phi_s,
\label{theta-den}
\end{equation}
where $\rho_s$ and $\phi_s=\rho_s v_1(D/2)$ are the number
density and packing density, respectively, of the saturated packing.
What is the thinnest possible covering associated with a saturated packing?
It immediately follows that the thinnest coverings among saturated congruent
sphere packings in $\mathbb{R}^d$ are
given by the saturated packings that have the minimal packing density $\phi_s^*$
in that space dimension and have covering density
\begin{equation}
\theta_s^*= 2^d \phi_s^*.
\end{equation}

We can also bound the packing density $\phi_s$ of any saturated sphere packing 
in $\mathbb{R}^d$ from above using upper bounds on the covering density.

\noindent{\bf Lemma 1:} {\sl The density $\phi_s$ of any saturated sphere packing 
in $\mathbb{R}^d$ is bounded from above according to}
\begin{equation}
 \phi_s \le \frac{5d}{2^d}+ \frac{d \ln(d)}{2^d}+ \frac{d\ln(\ln(d))}{2^d}.
\label{sat-up}
\end{equation}
The proof is trivial in light of the upper bound (\ref{cover-up}) on the covering
density and relation (\ref{theta-den}).

The standard random sequential addition (RSA) sphere packing
is a time-dependent process produced by randomly, irreversibly, 
and sequentially placing nonoverlapping spheres into a large region
${\cal V} \subset \mathbb{R}^d$ \cite{Wi66}. Initially, this large region
is empty of sphere centers and subsequently spheres are
added provided each attempted placement of a sphere does not overlap an existing
sphere in the packing. If an attempted placement results
in an overlap,  further attempts are made until the sphere
can be added to the packing at that time 
without violating the impenetrability constraint. 
For identical $d$-dimensional RSA spheres, the filling process
terminates at the {\it saturation limit} at infinitely long times
in the infinite-volume limit. Thus, in this limit, the RSA packing
is a saturated packing. The saturation density $\phi_s$ can only be
determined exactly in one dimension, where it is known
to be $\phi_s=0.747597\ldots$ \cite{Re63}. In higher dimensions,
the saturation density can only be determined from computer
simulations. Earlier work focused on two and three dimensions,
where is was found that  $\phi_s \approx 0.547$ 
\cite{Fe80} and $\phi_s\approx 0.38$  \cite{Co88}, respectively.
More recent numerical work has reported RSA saturation densities
for the first six space dimensions \cite{To06d}. It is important
to emphasize that upper bound (\ref{sat-up}) provides a very poor
estimate of the RSA saturation density; for example, for $d=3$,
it gives an upper bound greater than unity (2.32224, which is about 6.1
times larger than the value obtained from simulations) and for
$d=6$, it provides the upper bound $0.691402$, which is about
7.3 times larger than the value obtained from simulations \cite{To06d}.
Thus, bound (\ref{sat-up}) could only be realized by RSA saturated packings,
if at all, in high dimensions.

Therefore, it is useful here to compute the corresponding covering
densities for RSA packings. The numerical data for RSA saturation densities
for $2 \le d \le 6$ reported in Ref.~\cite{To06d}
was well approximated (with a correlation coefficient of 0.999) 
by the following form:
\begin{equation}
\phi_s= \frac{c_1}{2^d}+\frac{c_2 d}{2^d},
\label{linear}
\end{equation}
where $c_1=0.202048$ and $c_2=0.973872$. This can be used
to estimate the RSA saturation densities for $d\ge 7$.
However, even though the upper bound (\ref{sat-up}) on the packing density
of a saturated sphere packing grossly overestimates the RSA 
value for the first six space dimensions, we include  
in the fit function for $\phi_s$ a $d \ln(d)$ correction, which
will lead to a more conservative estimate of the covering density, as
explained shortly \cite{RSA}.
Including such a term, we find 
the following fitted function for the saturated RSA packing density: 
\begin{equation}
\phi_s=\frac{a_1}{2^d}+\frac{a_2 d}{2^d}+\frac{a_3 d \ln(d)}{2^d}
\label{log}
\end{equation}
with a correlation coefficient of $0.999$,
where $a_1=0.350648$, $a_2=0.87660$ and $a_3=0.041428$, does as 
well as (\ref{linear}) for $2 \le d \le 6$. Since (\ref{log}) predicts slightly higher
densities than (\ref{linear}) for $7 \le d \le 24$ \cite{fit}, we use it
to obtain the corresponding estimate of the RSA covering
density, namely,
\begin{equation}
\theta_s= a_1+ a_2 d+a_3 d \ln(d),
\label{log2}
\end{equation}
which is a slightly more conservative estimate, since (\ref{linear}) would
yield a slightly thinner covering density. 

In Table~\ref{RSA}, we provide
estimates of the RSA covering densities for selected dimensions
up through $d=24$. The  covering density for $d=1$ is determined from Re{\'n}yi's
exact saturation packing density value \cite{Re63} and multiplying it by 2. The values
for $2\le d \le 6$ is  are obtained from the reported
saturated density values in Ref.~\cite{To06d} and multiplying each
density value by $2^d$. The values reported for $d \ge 7$ are 
estimates obtained from the fitting formula (\ref{log2}).
Comparing Table~\ref{RSA} to Table~\ref{coverings} for the best known coverings,
we see that saturated RSA packings not only provide relatively thin coverings
but putatively represent the first non-lattices
that yield thinner coverings than the best known lattice coverings
beginning in about dimension 17. This suggests that saturated RSA packings
may be thinner than the previously best known coverings for $17 \le d \le 23$
and probably for some dimensions greater than 24.

\begin{table}
\caption{Covering density $\theta_s$ for  RSA packings
at the saturation state  in selected dimensions. The values
for the first 6 space dimensions are obtained from the reported
saturated density values given in Refs.~\cite{Re63,To06d} and 
using (\ref{theta-den}). The values reported for $d \ge 7$ are
estimates obtained from the fitting formula (\ref{log2}). Included
in the table are the corresponding the saturation packing densities. }
\label{RSA}
\renewcommand{\baselinestretch}{1.2} \small \normalsize
\centering
\begin{tabular} {|c|c|c|}
\multicolumn{3}{c}{~} \\\hline
 {Dimension, $d$} & Covering Density, $\theta_s$ & Packing Density, $\phi_s$ \\ \hline
1 &  1.4952& 0.74759\\

2 &  2.1880&  0.54700\\

3 &  3.0622&  0.38278 \\

4 &  4.0726&  0.25454\\

5 &  5.1526&  0.16102\\

6 &  6.0121&  0.09394\\
7 &  7.0512&  0.05508       \\
8 &  8.0526&  0.03145     \\
9 &  10.0706& 0.01769\\
10&  11.0860&  0.009834     \\
12&  12.1052&  0.002955 \\
16&  16.2141& $2.4740\times 10^{-4}$\\
17&  17.2482& $1.3159\times 10^{-4}$\\
18&  18.2848& $6.9751\times 10^{-5}$ \\
24&  24.5489& $1.4632\times 10^{-6}$
\\ \hline
\end{tabular}
\end{table}

\section{Results for the Quantizer Problem}
\label{results-quant}

Using the successive lower and upper bounds on the void exclusion probability
function $E_V(R)$ given in the previous section, we can, in principle, derive
corresponding bounds on the minimal error ${\cal G}_{min}$.
Moreover, one can obtain a variety of upper bounds on ${\cal G}_{min}$ using our approach by utilizing
the exact form of the void exclusion probability, when it is known, for
some point process at unit density. Since
a general point process must have an error  ${\cal G}$ that
is generally larger than the minimal ${\cal G}_{min}$, it trivially follows that
\begin{equation}
{\cal G}_{min} \le {\cal G}.
\label{upper}
\end{equation}

\subsection{Revisiting Zador's Bounds}
\label{Zador}

To illustrate how we can obtain bounds on ${\cal G}_{min}$ using our approach, we begin by rederiving
the following bounds due to Zador \cite{Za82}:
\begin{equation}
\frac{1}{(d+2)\pi}\Gamma(1+d/2)^{2/d} \le {\cal G}_{min} \le \frac{1}{d\pi}\Gamma(1+d/2)^{2/d}\Gamma(1+2/d).
\label{zador}
\end{equation}       
Consider the lower bound first. Combination of relation (\ref{G}) and lower bound
(\ref{EV-bound1}) yields at unit density
\begin{equation}
{\cal G}_{min} \ge \frac{2}{d} \int_0^{R_0} R[1 -v_1(R)] dR=\frac{1}{(d+2)\pi}\Gamma(1+d/2)^{2/d},
\end{equation}
which is seen to be equal to Zador's lower bound. Here $R_0=\Gamma(1+d/2)^{1/d}/\sqrt{\pi}$ 
is the zero of $1 - v_1(R)$. It is clear that a sphere of radius
$R_0$ has the smallest second moment of inertia of any solid $d$-dimensional
solid, and hence establishes the lower bound. 
The simplest example of a point process for which $E_V(R)$ is known
is the Poisson point process [cf. (\ref{Poisson})].
Substitution of (\ref{Poisson}) into (\ref{upper}) at unit density yields
\begin{equation}
{\cal G}_{min} \le \frac{2}{d} \int_0^\infty R \exp(-v_1(R)) dR=\frac{1}{d\pi}\Gamma(1+d/2)^{2/d}\Gamma(1+2/d),
\end{equation}
which is seen to be equal to Zador's upper bound. 
These derivations of the inequalities stated in (\ref{zador})
appear to be much simpler than the ones presented by Zador.

In the large-$d$ limit, Zador's upper and lower bounds become 
identical, and hence one obtains the exact asymptotic result
\begin{equation}
{\cal G}_{min} \rightarrow \frac{1}{2\pi e}=0.058550\ldots \quad \mbox{as}\quad d \rightarrow \infty.
\label{asympt}
\end{equation}
The convergence of Zador's bounds to the exact asymptotic limit is to
be contrasted with the sphere-packing problem in which the 
best upper and lower bounds on the maximal density become
exponentially far apart in the high-dimensional limit.

\subsection{Improved Upper Bounds}
\label{improve}

Improved upper bounds on ${\cal G}_{min}$ can be obtained by considering
those point processes corresponding to a sphere packing for which
the minimal pair separation is $D$ and lower bounds on the conditional function $G_V(R)$
for $R\ge D/2$. In what follows, we present two different upper bounds
on ${\cal G}_{min}$ based on this idea that improve upon Zador's upper bound.

For any packing of identical spheres with diameter $D$, the following exact
relations on the nearest-neighbor quantities apply for $R \le D/2$ \cite{To02a}:
\begin{equation}
E_V(R)=1 - 2^d \phi \left(\frac{R}{D}\right)^d, \quad H_V(R)=\frac{d 2^d \phi}{D} \left(\frac{R}{D}\right)^{d-1}, 
\quad G_V(R)=\frac{1}{ 1- 2^d \phi \left(\frac{R}{D}\right)^d}, \quad 0\le  R \le D/2.
\label{exact}
\end{equation}
Observe that $G_V(R)$ is a monotonically increasing function
of $R$ in the interval $[0,D/2]$.
From these equalities, it immediately follows that
\begin{equation}
E_V(D/2)=1 - \phi, \quad H_V(D/2)= \frac{2d \phi}{D}, 
\quad G_V(D/2)=\frac{1}{ 1- \phi}.
\label{equality}
\end{equation}

Consider now the class of sphere packings for which the
conditional nearest-neighbor function is bounded from below according to 
\begin{equation}
G_V(R) \ge \frac{1}{1-\phi} \qquad \mbox{for all} \quad R \ge D/2.
\label{GV-bound}
\end{equation}
This class of packings includes equilibrium (Gibbs) ensembles
of hard spheres along the disordered fluid branch of the
phase diagram \cite{Re59,To02a,To95a}, nonequilibrium disordered sphere
packings, such as the ``ghost" random sequential addition (RSA) process \cite{To06a},
and a large class of  lattice packings of spheres,
as will be described below. In the equilibrium cases, it is known that $G_V(R)$
is a monotonically increasing function of $R$ for $d\ge 2$ and 
thus using this property together with the equality $G_V(D/2)=1/(1-\phi)$
[cf. (\ref{equality})] means that the lower bound (\ref{GV-bound})
is obeyed. For one-dimensional equilibrium ``rods," the bound (\ref{GV-bound}) 
is sharp (exact) $G_V(R)$ for all realizable $\phi\in [0,1]$.
A bound of the type  (\ref{GV-bound}) was used to bound the related ``particle" mean
nearest-neighbor distance from above for different 
classes of sphere packings for all $d$ \cite{To95a}. 

Using definition (\ref{EV-GV}) and inequality (\ref{GV-bound}), the void exclusion probability function obeys  the following upper bound for $R\ge D/2$:
\begin{equation}
E_V(R) \le (1 -\phi)\exp\left\{ -\frac{2^d\phi}{1-\phi} \left[\left(\frac{R}{D}\right)^d -\frac{1}{2^d}\right]\right\}, \quad R \ge D/2.
\label{EV-1}
\end{equation}
Since any upper bound on the nonnegative function $E_V(R)$ leads to an upper bound on its first
moment, we then have upon use of (\ref{upper}), (\ref{exact}) and (\ref{EV-1}), the upper bound
\begin{equation}
{\cal G}_{min} \le \frac{4[\phi \Gamma(1+d/2)]^{2/d}}{d\pi}\left[\frac{(d+2(1-\phi))}{4(2+d)}
+\frac{(1-\phi)}{2d} \left(\frac{1-\phi}{ \phi}\right)^{2/d}\exp\left(\frac{\phi}{1-\phi}\right)
\Gamma\left(\frac{2}{d},\frac{\phi}{1-\phi}\right)\right],
\label{new-upper}
\end{equation}
where $\Gamma\left(s,x\right)\equiv \int_x^\infty t^{s-1} e^{-t} dt$
is the incomplete gamma function. Observe that the prefactor multiplying the bracketed
expression is $D^2/d$, where, in light of (\ref{v1}) and (\ref{den}), 
$D=2[\phi \Gamma(1+d/2)]^{1/d}/\sqrt{\pi}$,
assuming unit number density.
Note also that the upper bound (\ref{new-upper}) depends on a single
parameter, namely, the packing density $\phi$.
Thus, there is an optimal  packing density $\phi_{opt}\in [0,\phi_{max}]$
that yields the best (smallest) upper bound for any particular $d$,
where $\phi_{max}$ is the maximal packing density.
Since the right side of the inequality is a monotonically decreasing
function of $\phi$ for any $d$, then the optimal density $\phi_{opt}$
is, in principle, given by $\phi_{max}$.
It is noteworthy that the upper bound (\ref{new-upper}) for the optimal
choice $\phi=\phi_{max}$ may still be valid for a packing
even if the bound (\ref{EV-1}), upon which it is based, is
violated for $R$ of the order of $D$ because the exponential
tail can more than compensate for such a violation such that
the error [first moment of $E_V(R)$] is overestimated.  
Observe also that because 
\begin{equation}
\Gamma\left(\frac{2}{d},\frac{\phi}{1-\phi}\right)= \frac{1}{4}\left[ 
\left(\frac{1-\phi}{\phi}\right)^{2/d}\right] +{\cal O}(1) \quad (d \rightarrow \infty)
\end{equation}
the upper bound (\ref{new-upper}) tends to the exact asymptotic result
(\ref{asympt}) of $(2\phi e)^{-1}$.

Before discussing the optimal bounds, it is useful
to begin with an application of the upper bound (\ref{new-upper}) for the
{\it sub-optimal} 
case of a disordered sphere packing, namely, the aforementioned
{\it ghost} RSA packing process \cite{To06a}, which we now show
generally improves on Zador's upper bound. This represents the only
exactly solvable disordered sphere-packing model for
all realizable densities and in any dimension, as we now briefly 
describe.   The ghost RSA packing process involves a (time-dependent) sequential addition
of spheres in space subject to the nonoverlap condition. Not only 
is an attempted addition of a sphere rejected if it overlaps
an existing sphere of the packing, it is also rejected
if it overlaps any previously rejected sphere (called a 
``ghost" sphere). Unlike the standard RSA packing, the ghost
RSA packing does not become a saturated packing in the infinite-time limit.
All of the $n$-particle correlation functions for this nonequilibrium
model have been obtained analytically for any $d$, time $t$,
and for all realizable densities. For example, one can show that
the maximal density (achieved at infinite time) is given by
\begin{equation}
\phi=\frac{1}{2^d}
\end{equation}
and the associated pair correlation function is 
\begin{equation}
\rho^2g_2(r)= \frac{2\Theta(r-D)}{2-\alpha(r;D)},
\label{g2-GRSA}
\end{equation}
where $\Theta(x)$ is the unit step function,
equal to zero for $x<0$ and unity for $x \ge1$.
It is straightforward to verify that the upper bound 
on the exclusion probability $E_V(R)$ for this infinite-time
case obtained by using (\ref{g2-GRSA}) in the inequality (\ref{EV-bound2})
is always below the upper bound (\ref{EV-1}).
Therefore, the upper bound (\ref{new-upper}) is valid
at the maximal density, i.e., at $\phi=1/2^d$, we have
\begin{equation}
{\cal G}_{min} \le \frac{[ \Gamma(1+d/2)]^{2/d}}{d\pi}\left[\frac{(d+2(1-1/2^d))}{4(2+d)}
+\frac{2(1-1/2^d)^{(2+d)/d}}{d} \exp\left(\frac{1}{2^d-1}\right)
\Gamma\left(\frac{2}{d},\frac{1}{2^d-1}\right)\right].
\label{upper-GRSA}
\end{equation}
For $d=1$, 2 and 3, this upper bound yields $0.166666\dots$, 
$0.124339\ldots$ and $0.106797\ldots$, respectively, which is
to be compared to Zador's upper bound, which gives
$0.5$, $0.159154\dots$ and $0.115802\ldots$, respectively.
We note that in the large-$d$ limit, the upper bound (\ref{upper-GRSA}) yields
the exact asymptotic result (\ref{asympt}), which implies
that the upper bound (\ref{EV-1}) on $E_V(R)$ becomes exact for ghost RSA packings, 
tending to the unit step function in this asymptotic limit, i.e.,
\begin{equation}
E_V(R) \rightarrow \Theta(r-D) \qquad (d \rightarrow \infty).
\end{equation}
This asymptotic result implies the following corresponding one for
the void nearest-neighbor probability density function:
\begin{equation}
H_V(R) \rightarrow \delta(r-D) \qquad (d \rightarrow \infty).
\end{equation}

\begin{figure}[bthp]
\centerline{ \includegraphics[width=3.5in]{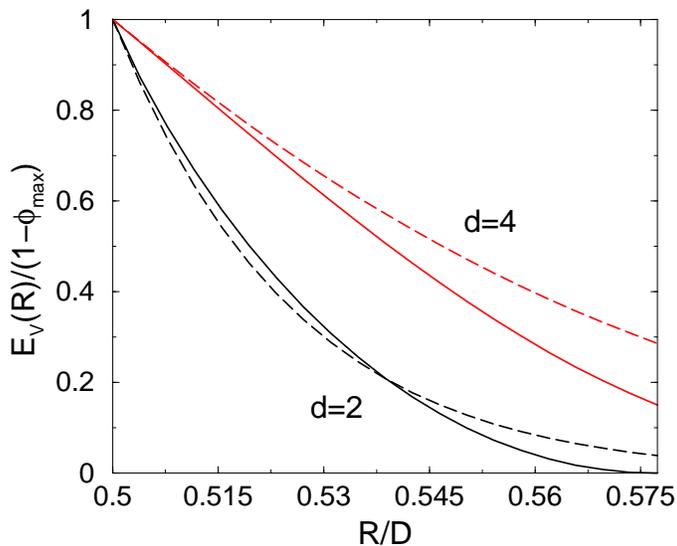}}
\caption{ (Color online) Comparison of the exact result
for $E_V$ scaled by $1-\phi_{max}$ for the optimal
lattice packings for $d=2$ ($A_2^*$) and $d=4$ ($D_4$), 
as obtained from (\ref{EV-packing2}), for $1/2 \le R/D \le 1/\sqrt{3}$
(solid curves) to the corresponding estimates obtained from
(\ref{EV-1}) for these cases (dashed curves).
It is only for the case $d=2$ that estimate (\ref{EV-1}) is not
a rigorous pointwise upper bound on the exact void exclusion
probability for $1 \le d \le 24$ and,   likely, for 
 $d >24$. The exponential tail associated with (\ref{EV-1}) 
more than compensates for the
narrow pointwise violation for the special case $d=2$, resulting in a strict upper bound on the first moment of $E_V(R)$, i.e., (\ref{new-upper}) remains a strict upper bound for $d=2$. }
\label{Ev-bounds}
\end{figure}

We now return to finding the optimal (smallest) upper bound (\ref{new-upper}) for each dimension.
For $d=1$, the optimal density is $\phi_{opt}=\phi_{max}=1$, which produces
the sharp bound
\begin{equation}
{\cal G}_{min} \le \frac{1}{12}=0.083333\ldots.
\end{equation}
This bound is exact in this case because the inequality (\ref{GV-bound}) 
is exact for all realizable densities for  equilibrium hard ``rods,"
including at $\phi=1$, which corresponds to the optimal integer lattice packing.
This is to be contrasted with Zador's upper bound, which yields
1/2 for $d=1$ and is far from the exact result. The improved
upper bound (\ref{new-upper}) in the higher dimensions
reported in Table \ref{Tab-eta-3} is obtained by evaluating it
at the densities of the densest known lattice packings in these respective
dimensions \cite{Co93}. We note that it is only for optimal
triangular lattice packing in $\mathbb{R}^2$  that the upper bound (\ref{EV-1})
on $E_V(R)$ is violated pointwise for a small range of $R$ around $R/D=1/2$ 
[inequality (\ref{EV-1}) is obeyed for $R/D \ge 0.539$ and in the vicinity of $R/D=0.5$], but the exponential tail associated with  (\ref{EV-1}) more than compensates for this 
narrow pointwise violation, resulting in a strict upper bound 
 on the first moment of $E_V(R)$, i.e.,
(\ref{new-upper}) remains a strict upper bound for $d=2$.
 Using relation (\ref{EV-packing2}) and lattice
coordination properties, it is easily verified that the (\ref{EV-1}) is a strict upper bound
on the void exclusion probability for the densest known lattice packings for 
all $R\ge D/2$ and $3 \le d \le 24$ as well as $d=1$, and hence inequality (\ref{new-upper})
provides a strict upper bound on the scaled error for all of these lattices. 
For illustration purposes, we compare in Fig.~\ref{Ev-bounds} the exact result
for $E_V$ obtained from (\ref{EV-packing2}) to the estimate
(\ref{EV-1}) for the cases $d=2$ and $d=4$ for $1/2 \le R/D \le 1/\sqrt{3}$.
The upper bound
(\ref{new-upper}) is generally appreciably tighter than Zador's upper bound
for low to moderately high dimensions.

\begin{table}
\renewcommand{\baselinestretch}{1.2} \small \normalsize
\centering
\caption{Comparison of the best known quantizers in selected dimensions to the conjectured
lower bound due to Conway and Sloane and the improved upper bound (\ref{new-upper}). }
\begin{tabular} {|c|c|c|c|c|}
\multicolumn{5}{c}{~} \\\hline
$d$ & Quantizer & Scaled Error, $\cal G$ & Conjectured  & Improved  \\ 
 &  &  & Lower bound & Upper Bound \\ \hline
1  & $A_1^*= \mathbb{Z}$  & 0.083333   & 0.083333   & 0.083333 \\
2  & $A_2^*\equiv A_2$    & 0.080188 & 0.080188& 0.080267 \\
3  & $A_3^*\equiv D_3^*$  & 0.078543 & 0.077875& 0.079724 \\
4  & $D_4^*\equiv D_4$  & 0.076603& 0.07609 & 0.078823 \\
5  & $\Lambda_5^{2*}$ & 0.075625 & 0.07465 & 0.078731\\
6  & $E_6^*$  & 0.074244 & 0.07347 & 0.077779 \\
7  & $\Lambda_7^{3*}$ & 0.073116 & 0.07248  & 0.076858\\
8  & $E_8^*= E_8$  & 0.071682 & 0.07163      & 0.075654\\
9 &  $L^{AE}_9$ & 0.071626 & 0.070902 &  0.075552\\ 
10&  $D^+_{10}$ & 0.070814  & 0.070405  & 0.074856\\ 
12  & $K_{12}^*\equiv K_{12}$ & 0.070100 &  0.06918     & 0.073185\\
16 & $\Lambda_{16}^*\equiv \Lambda_{16}$  & 0.068299 & 0.06759      & 0.070399\\
24 & $\Lambda_{24}^*=\Lambda_{24}$  & 0.065771 & 0.06561     & 0.067209 \\ \hline
\end{tabular}

\label{Tab-eta-3}
\end{table}

\subsection{Results for Saturated Packings}
\label{sat-quant}

\begin{figure}
 \centerline{ \includegraphics[width=3.0in,keepaspectratio,clip]{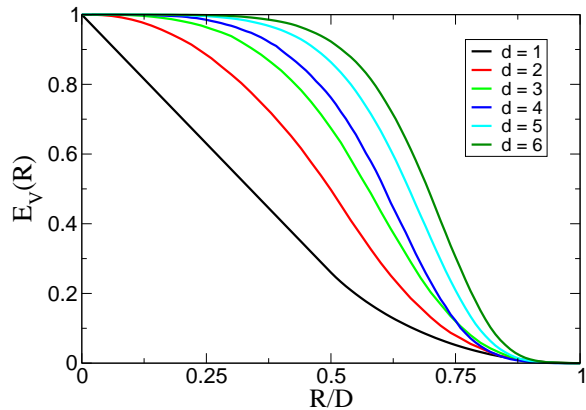}}
\caption{ (Color online) The void exclusion probability $E_V(R)$ for saturated RSA packings
of congruent spheres of diameter $D$ for the first six space dimensions.}
\label{RSA-all}
\end{figure}

For any saturated packing of identical spheres
of diameter $D$, $E_V(R)$ by definition is exactly zero 
for $R$ beyond the diameter, i.e.,
\begin{equation}
E_V(R)=0 \quad R \ge D.
\end{equation}
In the particular case of saturated RSA packings, the void
exclusion probability can computed using the same techniques
described in Ref.~\cite{To06d} for the first six space dimensions.
These results are summarized in Fig.~\ref{RSA-all}. The corresponding
quantizer errors for these dimensions are listed in Table~\ref{RSA-quant}.
We see that the discrepancies between the saturated RSA quantizer
error improves as $d$ increases as compared to the best known
quantizer error reported in Table~\ref{quant}.

\begin{table}
\caption{The quantizer errors for saturated RSA packings in the first
six space dimensions. We denote by ${\cal G}_s$ the quantizer error for
such as saturated packing. }
\label{RSA-quant}
\renewcommand{\baselinestretch}{1.2} \small \normalsize 
\centering
\begin{tabular} {|c|c|}
\multicolumn{2}{c}{~} \\\hline
 {Dimension, $d$} & Quantizer Error, ${\cal G}_s$  \\ \hline
1 &  0.11558\\

2 &  0.09900\\

3 &  0.09232\\

4 &  0.08410\\

5 &  0.07960\\

6 &  0.07799
\\ \hline 
\end{tabular} 
\end{table}

\noindent{\bf Lemma 2:} {\sl Saturated
sphere packings in $\mathbb{R}^d$ possess void nearest-neighbor
functions that tend to the following high-dimensional
asymptotic behaviors:}
\begin{equation}
E_V(R) \rightarrow \Theta(r-D), \qquad H_V(R) \rightarrow \delta(r-D)\quad (d \rightarrow \infty).
\label{lemma2}
\end{equation}

For any saturated packing at packing density $\phi_s$, it is clear
that $E_V(R)$ is bounded from above for $R \ge D/2$ as follows:
\begin{equation}
E_V(R) \le 1- \phi_s, \quad D/2 \le R \le D.
\label{RSA-bound}
\end{equation}
Let ${\cal G}_s$ denote the scaled dimensionless quantizer error
for a saturated packing. Combination of the expression for $E_V(R)$ in (\ref{exact}) 
and (\ref{RSA-bound}) yields the following upper bound on ${\cal G}_s$:
\begin{equation}
{\cal G}_s \le \frac{4[\phi_s \Gamma(1+d/2)]^{2/d}}{d\pi}\left[\frac{(d+2(1-\phi_s))}{4(2+d)}
+\frac{3(1-\phi_s)}{8}
\right],
\label{new-upper2}
\end{equation}
The fact that this upper bound becomes exact in the high-dimensional limit
(that is, it tends to $(2\pi e)^{-1}$), implies that $E_V(R)$ and $H_V(R)$ for 
a saturated packing tends to the unit step function and radial delta function,
as specified by (\ref{lemma2}). Not surprisingly, the bound (\ref{new-upper2})
is not that tight in relatively low dimensions.

\section{Concluding Remarks and Discussion}
\label{conc}

We have reformulated the covering and quantizer 
problems as the determination of the ground states of interacting point 
particles in   $\mathbb{R}^d$ that generally involve single-body, two-body,
three-body, and higher-body interactions; see Sec.~\ref{reform}. The $n$-body interaction
is directly related to a purely geometrical problem, namely,
the intersection volume of $n$ spheres of radius $R$ centered at
$n$ arbitrary points of the system in $\mathbb{R}^d$.   
This was done by linking  
the covering and quantizer problems to certain 
optimization problems involving the  ``void" nearest-neighbor functions.
This reformulation 
allows one to employ  theoretical and numerical optimization techniques 
to solve these energy minimization problems. 

A key finding is 
 that disordered saturated sphere packings provide relatively thin coverings
and may yield thinner coverings than the best known lattice coverings
in sufficiently large dimensions. In the case of the quantizer problem, 
we derived improved upper bounds on  the quantizer error that utilize
sphere-packing solutions. These improved bounds are generally substantially 
sharper than an existing upper bound due to Zador in low to moderately large dimensions. Moreover, we showed that disordered  saturated sphere packings yield
relatively good quantizers. 
Our reformulation helps to explain why the known solutions of quantizer and covering problems
are identical in the first three space dimensions and why they can be 
different for $d \ge 4$. In the first three space dimensions, the best known solutions of the sphere packing and number variance problems are directly related to those of the covering and quantizer problems, but such relationships 
may or may not  exist for $d \ge 4$, depending on the peculiarities
of the dimensions involved. It is clear that as $d$ becomes large,
the quantizer problem becomes the easiest to solve among the
four ground-state problems considered in this paper, since,
unlike the other three problems,
the asymptotic quantizer error tends to the same limit
independent of the configuration of the point process.

The detection of gravitational waves from various astrophysical
sources has and will be searched for in the output of
interferometric networks \cite{LL} by correlating the noisy output of
each interferometer with a set of theoretical waveform templates \cite{Pr07}.
Depending upon the source, the parameter space is generally multidimenisonal
and can be as large as $d=17$ or larger for inspiraling  binary black holes.
The templates must {\it cover} the space and the challenge is 
to place them in some optimal fashion such that the fewest templates are
used to reduce computational cost without reducing the detectability
of the signals. 
  
Optimal template placement for gravitational wave data analysis has proved to be 
highly nontrivial. One solution proposed for the optimal placement in 
flat (Euclidean) space is to simply use the optimal solution to the
covering problem \cite{Pr07}. However, this requires every point
in the parameter space to be covered by at least one template, 
which rapidly becomes inefficient in higher dimensions when
optimal lattice covering solutions are employed. 
Another approach to template placement  consists in relaxing the strict requirement of
complete coverage for a given mismatch, and instead require 
coverage only with a certain confidence \cite{Pr07}.
Such stochastic approaches have involved randomly placing 
templates in the parameter space, accompanied by a ``pruning" step in which 
``redundant" templates, which are deemed to lie too close to each other, are removed \cite{Ha08}.
The pruning step may be a complication that can be avoided, as discussed below.
Another approach is to place spherical  templates down according to
a Poisson point process \cite{Me09}, which has been claimed to provide good solutions for $d >10$.
 The problem with the latter approach 
is that there will be numerous multiple overlaps of templates, which
only increases as the number density of templates (intensity
of the Poisson point process) is increased in order to 
cover as much of the space as is computationally feasible.

The results of the present study suggest alternative solutions to the
optimal template construction problem. First, we remark
that if the covering of space by the templates is relaxed,
then it is possible that the optimal lattice quantizers
could serve as good solutions  
in relatively low dimensions ($d \le 10$)
because the mean square error is minimized. Second, 
in such relatively low dimensions, we have shown that saturated RSA
sphere packings provide both relatively good coverings and quantizers,
and hence may be useful template-based constructions for the search of gravitational waves.
Indeed, we have shown that saturated RSA sphere packings are expected to become better solutions as
$d$ becomes large. However, for $d \ge 10$, it will be computationally costly
to create truly saturated packings, which by definition provide coverings of
 space (see Sec.~\ref{results-cover}).
However, the existing stochastic approaches do not require
complete coverage of space and hence an {\it unsaturated} RSA 
sphere packing that gets relatively close to the saturation state
might still be more computationally efficient than either the random placement/pruning
technique \cite{Ha08} (because pruning is unnecessary) or the Poisson placement procedure
\cite{Me09} (because far fewer spheres need to be added).
Moreover, when the template parameter space is curved, which occurs in practice \cite{Me09},
the RSA sphere packing would be computational faster to adapt
than lattice solutions.

In future work, we will explore whether our reformulations of the covering and
quantizer problems as ground-state problems of many-body interactions
of the form (\ref{full}) can facilitate the search for better solutions
to these optimization tasks. Clearly, a computational challenge in
high dimensions will be the determination of the intersection volume $ v_n^{\mbox{\scriptsize int}}$
of $n$ spheres of radius $R$ at $n$ different locations in $\mathbb{R}^d$ for
sufficiently large $n$.
However, it is possible that that the series representations (\ref{EV}) for $E_V(R)$
and bounds on this quantity [cf. Sec.~\ref{bounds-E}] can be used to devise
useful approximations of the monotonic function $E_V(R)$, which should
be zero for $R \ge 2{\cal R}_c$, where ${\cal R}_c$ is the bounded covering radius.
Such approximation could be employed to evolve an initial guess of the configuration
of points within a fundamental cell to useful but sub-optimal solutions, which
upon further refinement could suggest novel solutions. In short, the implications
of our reformulations to discover better solutions to the covering and
quantizer problems in selected dimensions have yet to be fully investigated 
and deserves future attention.

\begin{acknowledgments}
I am very grateful to Yang Jiao and Chase Zachary for their
assistance in creating many of the figures for this manuscript
and for many useful discussions. 
I thank Andreas Str{\" o}mbergsson  and Zeev Rudnick for 
their generous help in providing efficient algorithms to compute
the Epstein zeta function for $d=12, 16$ and 24. I also thank
Henry Cohn for introducing me to the  quantizer
problem as well as for many helpful discussions. This work was supported by the
Office of Basic Energy Sciences, U.S. Department of Energy,
under Grant No. DE-FG02-04-ER46108. 
\end{acknowledgments}

\appendix

\section{Computing the Number Variance via the Epstein Zeta Function for $d=12,16$ and $24$}
\label{num-Ep}
       
Here we summarize the steps in computing the 
asymptotic number-variance coefficient (\ref{num})
for the lattices $K_{12}$, $\Lambda_{16}$ and $\Lambda_{24}$,
which correspond to the densest lattice packings in dimensions
$12,16$ and 24, respectively. Importantly, the sum  in (\ref{num})
converges slowly. We noted in Sec.~\ref{num-var} that the {\it dual} of the lattice
that minimizes the Epstein zeta function $Z_{\Lambda}(s)$ [defined
by (\ref{epstein})] at $s=(d+1)/2$ among all lattices will minimize the asymptotic
number-variance coefficient (\ref{num}) among lattices.
We will exploit number-theoretic representations
of the Epstein zeta function that enable its efficient
numerical evaluation and thus efficient computation
of the asymptotic number-variance coefficient (\ref{num})
using the aforementioned duality relation.

First, let us note that Epstein zeta function $Z_{\Lambda}(s)$ 
(\ref{epstein}) can be rewritten as follows:
\begin{equation}
Z_{\Lambda}(s)= \sum_{i=1} \frac{Z_i}{p_i^{2s}},
\label{zeta}
\end{equation}
where $Z_i$ is the coordination number at a radial distance
$p_i$ from some lattice point in the lattice $\Lambda$.
[Note that the Epstein zeta function defined in this
way applies to a general periodic point process provided
that $Z_i$ is interpreted in the generalized sense discussed
in (\ref{def}).] The quantities $Z_i$ and $p_i$ 
for many well-known lattices in $\mathbb{R}^d$
can be obtained analytically
using the \emph{theta series} for a lattice $\Lambda$, which is defined by
\begin{eqnarray}
\Theta_{\Lambda}(p) = 1+\sum_{i=1}^{\infty} Z_i q^{p_i^2},
\end{eqnarray}
and is directly related to the quadratic form
associated with the lattice \cite{Co93}. 
This series expression can usually be generated from
the simpler functions $\theta_2,$ $\theta_3$, and $\theta_4$, which are defined by \cite{Co93}:
\begin{eqnarray}
\theta_2(q) &= 2\sum_{m=0}^{+\infty} q^{(m+1/2)^2}\\
\theta_3(q) &= 1+2\sum_{m=1}^{+\infty} q^{m^2}\\
\theta_4(q) &= 1+2\sum_{m=1}^{+\infty} (-q)^{m^2}.
\end{eqnarray}
Specifically, for the  $K_{12}$, $\Lambda_{16}$ and $\Lambda_{24}$ lattices,
the associated theta series are given by \cite{Co93}
\begin{eqnarray}
\Theta_{K_{12}} &=& \phi_0(2q)^6+45 \phi_0(2q)^2\phi_1(2q)^4+18\phi_1(2q)^6\nonumber\\
&=& 1+756 q^4+ 4032q^6+20412q^8+ \cdots 
\end{eqnarray}
\begin{eqnarray}
\Theta_{\Lambda_{16}} &=& \frac{1}{2}[\theta_2(2q)^{16}+ \theta_3(2q)^{16}
+ \theta_4(2q)^{16}+ 30\theta_2(2q)^{8} \theta_3(2q)^{8}]\nonumber\\
&=& 1+4320q^4+ 61440q^6+ \cdots 
\end{eqnarray}
\begin{eqnarray}
\Theta_{\Lambda_{24}} &=& \frac{1}{8}[\theta_2(q)^{8}+ \theta_3(q)^{8}
+ \theta_4(q)^{8}]^3
- \frac{45}{16}[\theta_2(q)\theta_3(q)\theta_4(q)]^8\nonumber\\
&=& 1+196560q^4+ 16773120q^6+ \cdots 
\end{eqnarray}
where 
\begin{eqnarray}
\phi_0&=&\theta_2(2q)\theta_2(6q)+ \theta_3(2q)\theta_3(6q), \\
\phi_1&=&\theta_2(2q)\theta_3(6q)+ \theta_3(2q)\theta_2(6q) .
\end{eqnarray}

Direct evaluation
of (\ref{zeta}) has the same convergence problems that the direct evaluation of the asymptotic number-variance coefficient (\ref{num}).
However, we can exploit alternative number-theoretic representations
of (\ref{zeta}) to facilitate its evaluation. In particular,
there is an expression for the Epstein zeta function that
can be derived using Poisson summation \cite{Sa06}:
\begin{eqnarray}
F_{\Lambda}(s)&=&\pi^{-s}\Gamma(s) Z_{\Lambda}(s) \nonumber\\
&=& \frac{1}{s-\frac{d}{2}} -\frac{1}{s} +  \sum_
{\stackrel{ \displaystyle  {\bf p} \in \Lambda }
{ \displaystyle {\bf p} \neq {\bf 0}  }}
G(s,\pi |{\bf p}|^2)+ \sum_{\stackrel{ \displaystyle  {\bf p} \in  \Lambda^* }
{  \displaystyle {\bf p} \neq {\bf 0}  }}
G(\frac{d}{2}-s,\pi |{\bf p}|^2),
\label{F}
\end{eqnarray}
where
\begin{equation}
G(s,x)=x^{-s}\Gamma(s,x)
\end{equation}
and $\Gamma(s,x)=\int_x^\infty e^{-t} t^{s-1} dt$ is the
complementary incomplete Gamma function. It is important
to note that the volumes  of the fundamental cells of the lattice and its dual
associated with the first and second sums in (\ref{F}), respectively,
are both taken to be unity here. Using the appropriate theta
series given above for the lattices corresponding to the
densest lattice packings for $d=12,16$ and 24 and expression
(\ref{F}) for $s=(d+1)/2$, one finds the corresponding
Epstein zeta functions to be
\begin{eqnarray}
Z_{K_{12}}(s=13/2)&=&   12.527470092112\ldots \label{12}\\
Z_{\Lambda_{16}}(s=17/2)&=&2.606378060701\ldots \label{16}\\
Z_{\Lambda_{24}}(s=25/2)&=& 0.026464258871\ldots \label{24}
\end{eqnarray}
Now since all of these lattices are self-dual (i.e.,
$K_{12}\equiv K^*_{12}$, $\Lambda_{16}\equiv \Lambda^*_{16}$, 
$\Lambda_{24}\equiv \Lambda^*_{24}$), we can directly determine
the corresponding asymptotic number-variance from
(\ref{num}) by replacing the sum therein with
the appropriate evaluation of the Epstein zeta
function  specified by relations (\ref{12})-(\ref{24}).
The asymptotic number-variance values for $d=12,16$ and 24 
reported in Table~\ref{var} were obtained in this fashion.

\section{Intersection Volume of Three Spheres in Three Dimensions}
\label{v3-int}

For $d=3$, the intersection volume $v_3^{\mbox{\scriptsize int}}(r,s,t;R)$ of three identical
spheres of radius $R$ whose centers are separated by the distances
$r$, $s$, and $t$ for $R > R_T$  is given by \cite{Po64} 
\begin{eqnarray}
v^{\mbox{\scriptsize int}}_3 (r,s,t;R) &=& \frac{Q}{6} rst +\frac{4}{3} \tan^{-1}\Big(\frac{Q \cdot rst}{r^2+s^2+t^2-8R^2}
\Big) \nonumber \\
&& -
r(R^2-r^2/12) \tan^{-1}\Big(\frac{2Q st}{-r^2+s^2+t^2}\Big)  \nonumber \\
&& - 
s(R^2-s^2/12) \tan^{-1}\Big(\frac{2Q  rt}{r^2-s^2+t^2}\Big)  \nonumber \\
&& -
t(R^2-t^2/12) \tan^{-1}\Big(\frac{2Q  rs}{r^2+s^2-t^2}\Big),  
\label{v3int}
\end{eqnarray}
where $0 \le \tan^{-1}x \le \pi$,
\begin{equation}
R_T=\frac{rst}{[(r+s+t)(-r+s+t)(r-s+t)(r+s-t)]^{1/2}},
\label{circum}
\end{equation}
is the circumradius of the triangle with side length lengths $r$, $s$, and  $t$
and $Q=\sqrt{R^2-R_T^2}/{R_T}$.


\begin{thebibliography}{30}
\bibitem{support}
The support of a function is the set of points where the function is not zero.
A function has compact support if it is zero outside of a compact set.
\bibitem{duality}
Mathematicians usually define a dual Bravais lattice to have a
fundamental cell volume $v_{F^*} = 1/v_{F}$ (i.e., without the factor of
$(2\pi)^d$), in which case self-duality is defined with respect to unit
density; see Ref.~\cite{Co93}.
\bibitem{laminate}
A laminated lattice $\Lambda_d$ in $\mathbb{R}^d$ is built up of layers of 
$(d-1)$-dimensional lattices in $d$-dimensional Euclidean space. 
Note that a general lattice has been denoted by $\Lambda$,
which should not be confused with the specific laminated
lattice $\Lambda_d$. These are common notations for
both objects and therefore we adhere to this convention.
\bibitem{hv}
For $d=1$, the condition on $H_V(0)$ in Eq.~\ref{origin} should
be replaced by $H_V(0)=2\rho$.
\bibitem{ground}
There is an uncountably infinite number of periodic
point configurations in $\mathbb{R}^d$ in which circumscribed overlapping
spheres of radius ${\cal R}_c$  around each of the points just cover the 
space and renders $E_V({\cal R}_c)=0$ such that for  $R> {\cal R}_c$.
The point configuration that minimizes the support
of $E_V(R)$, i.e., minimizes ${\cal R}_c$ (called ${\cal R}_c^{min}$) is the 
unique ground-state  configuration or the optimal covering.
Twice the covering radius $2{\cal R}_c$ for a particular point process determines
the ``effective range" of the interaction associated with $E_V(R)$; see 
the text for further explanation.
\bibitem{RSA}
Note that leading order term in either (\ref{linear}) or (\ref{log})
involving $1/2^d$ is a dominant dimensional contribution
for RSA saturation densities in relatively low dimensions for good theoretical
reasons \cite{To06d}. The fact that this term does not appear
in the upper bound (\ref{sat-up}) is another reason 
why it could only be realizable by RSA saturated packings
in high dimensions. If true, then the $d \ln (d)$ and $d \ln(\ln(d))$
terms in (\ref{sat-up}) are high-dimensional asymptotic corrections.
\bibitem{fit}
Observe that the extrapolation of the conservative fit function (\ref{log}) 
to $d=24$ gives a density that is about 3.7 percent larger than that predicted
by (\ref{linear}). This percentage difference between the two fit functions 
decreases as $d$ decreases to $d=7$.
\end{thebibliography}

\end{document}